\documentclass[journal]{IEEEtran}
\usepackage{amsmath,amssymb,dsfont,graphicx, tikz}
\usepackage{amsthm}
\usepackage{calc}

\title{Invertible Extractors and Wiretap Protocols} \author{Mahdi Cheraghchi,
  \IEEEmembership{Member, IEEE}, Fr\'ed\'eric Didier,
  and \\ Amin Shokrollahi,
  \IEEEmembership{Fellow, IEEE}

  \thanks{{\footnotesize 
  M.~Cheraghchi is with the Department of Computer Science, University of Texas at Austin, USA (email: mahdi@cs.utexas.edu).  F.~Didier is 
      with Google, Inc.\ (email: frederic.didier@gmail.com). 
  A.~Shokrollahi is with the School of Computer and Communication
      Sciences, Ecole Polytechnique F\'{e}d\'{e}rale de Lausanne
      (EPFL), Switzerland (email: amin.shokrollahi@epfl.ch). Part of the work 
      was done while the first two authors were with the Laboratory of Algorithms (ALGO)
      at EPFL. This work was supported
      by the Swiss NSF grant 200020-115983/1 and the ERC Advanced investigator grant 228021. A preliminary
      summary of this work appears (under the same title) in
      proceedings of the 2009 IEEE International Symposium on
      Information Theory.}}  }

\date{}

\newtheorem{thm}{Theorem}
\newtheorem{coro}[thm]{Corollary}
\newtheorem{lem}[thm]{Lemma}
\newtheorem{prop}[thm]{Proposition}
\newtheorem*{claims}{Claim}

\theoremstyle{definition}
\newtheorem{rem}[thm]{Remark}
\newtheorem{defn}[thm]{Definition}

\newcommand{\rv}[1]{{#1}}
\newcommand{\F}{\ensuremath{\mathds{F}}}
\newcommand{\N}{\mathds{N}}
\newcommand{\eps}{\epsilon}
\renewcommand{\varepsilon}{\epsilon}
\newcommand{\veps}{\epsilon}

\newcommand{\U}{\mathcal{U}}
\newcommand{\C}{\mathcal{C}}
\newcommand{\Z}{\mathds{Z}}

\newcommand{\cC}{\mathcal{C}}

\newcommand{\eqdef}{:=}

\newcommand{\cX}{\mathcal{X}}
\newcommand{\cY}{\mathcal{Y}}

\newcommand{\cE}{\mathcal{E}}

\newcommand{\zo}{\{0,1\}}

\newcommand{\dist}{{\mathsf{dist}}}
\newcommand{\rdist}{{\mathsf{rdist}}}
\newcommand{\rk}{{\mathsf{rank}}}
\newcommand{\poly}{{\mathsf{poly}}}

\newcommand{\Exp}{\mathbb{E}}

\newcommand{\bou}{\mathsf{AExt}}
\newcommand{\kz}{\mathsf{SFExt}}
\newcommand{\extr}{\mathsf{Ext}}

\newcommand{\TRASH}[1]{}

\providecommand{\eqref}[1]{(\ref{#1})}

\renewcommand{\lg}{\log}

\newenvironment{Proof}{\begin{proof}}{\end{proof}}

\newcommand{\Tk}{{\tilde{k}}}

\renewcommand{\qed}{\hfill \IEEEQED}

\usepackage[
     plainpages=false,
     pdftitle={Invertible Extractors and Wiretap Protocols},
     pdfsubject={Invertible Extractors and Wiretap Protocols},
     pdfstartview=FitH,
     bookmarks=true,
     colorlinks=false,
     bookmarksopen=false,
     bookmarksnumbered=true,
     pdfhighlight=/I,
     hyperfigures=true,
     pdfborder=0 0 0,
     pdfauthor={Mahdi Cheraghchi}]{hyperref}

\providecommand{\cites}[1]{\cite{#1}}
\date{}

\begin{document}

\maketitle

\begin{abstract}
  A wiretap protocol is a pair of randomized encoding and decoding
  functions such that knowledge of a bounded fraction of the encoding
  of a message reveals essentially no information about the message,
  while knowledge of the entire encoding reveals the message using the
  decoder.  In this paper we study the notion of efficiently invertible
  extractors and show that a wiretap protocol can be constructed from
  such an extractor.  We will then construct invertible extractors for
  symbol-fixing, affine, and general sources and apply them to create
  wiretap protocols with asymptotically optimal trade-offs between
  their rate (ratio of the length of the message versus its encoding)
  and resilience (ratio of the observed positions of the encoding and
  the length of the encoding).  We will then apply our results to
  create wiretap protocols for challenging communication problems,
  such as active intruders who change portions of the encoding,
  network coding, and intruders observing arbitrary boolean functions
  of the encoding. \\
  As a by-product of our constructions we obtain new explicit 
  extractors for a restricted family of affine sources over large fields
  (that in particular generalizes the notion of symbol-fixing sources) 
  which is of independent interest. These extractors are able to 
  extract the entire source entropy with zero error.

\end{abstract}

\section{Introduction} \label{sec:intro}

\IEEEPARstart{S}{uppose} that Alice wants to send a message to Bob through a
communication channel, and that the message is \emph{partially}
observable by an intruder.  This scenario arises in various practical
situations.  For instance, in a packet network, the sequence
transmitted by Alice through the channel can be fragmented into small
packets at the source and/or along the way. Then, different packets might
be routed through different paths in the network in which an intruder
may have compromised some of the intermediate routers.  An example
that is similar in spirit is furnished by transmission of a piece of
information from multiple senders to one receiver, across different
delivery media, such as satellite, wireless, and/or wired networks.
Due to limited resources, a potential intruder may be able to observe
only a fraction of the lines of transmission, and hence only partially
observe the message.
As another example, one can consider secure storage of data on a
distributed medium that is physically accessible in parts by an
intruder, or a sensitive file on a hard drive that is erased from the
file system but is only partially overwritten with new or random
information, and hence, is partially exposed to a malicious party.

An obvious approach to solve this problem is to use a secret key to
encrypt the information at the source. However, almost all practical
cryptographic techniques are shown to be secure only under unproven
hardness assumptions and the assumption that the intruder possesses
bounded computational power.  This might be undesirable in certain
situations. Moreover, the key agreement problem has its own
challenges.

In this paper, we assume the
intruder to be \emph{information theoretically} limited, and our goal
will be to employ this limitation and construct a protocol that
provides unconditional, infor\-mation-theoretic security, even in the
presence of a computationally unbounded adversary.

The problem described above was first formalized by Wyner
\cite{ref:Wyner1} and subsequently by Ozarow and Wyner
\cite{ref:Wyner2} as an infor\-mation-theoretic problem.  In its most
basic setting, this problem is known as the \emph{wiretap~II
  problem}\index{wiretap~II problem} (the description given here
follows from \cite{ref:Wyner2}):
  
\begin{quote}
Consider a communication system with a {source} which outputs a
sequence $X = (X_1,\ldots, X_m)$ in $\zo^m$ uniformly at random.  A
randomized algorithm, called the encoder,
maps the output of the source to a binary string $Y \in \zo^n$.  The
output of the encoder is then sent through a noiseless channel (called
\emph{the direct channel}\index{direct channel}) and is eventually
delivered to a decoder\footnote{Ozarow and Wyner also consider the
  case in which the decoder errs with negligible probability, but we
  are going to consider only error-free decoders.} $D$ which maps $Y$
back to $X$.
Along the way, an intruder arbitrarily picks a subset $S \subseteq
[n] := \{1,\ldots,n\}$ of size $t \leq n$, and is allowed to observe\footnote{ For a
  vector $x = (x_1, x_2, \ldots, x_n)$ and a subset $S \subseteq [n]$,
  we denote by $x|_S$ the vector of length $|S|$ that is obtained from
  $x$ by removing all the coordinates $x_i$, $i \notin S$.  } $W :=
Y|_S$ (through a so-called \emph{wiretap channel}\index{wiretap
  channel}), i.e., $Y$ on the coordinate positions corresponding to
the set $S$.
The goal is to make sure that the intruder learns as little as
possible about $X$, regardless of the choice of $S$.
\end{quote}

\begin{figure*}
\centerline{\includegraphics[width=0.9\textwidth]{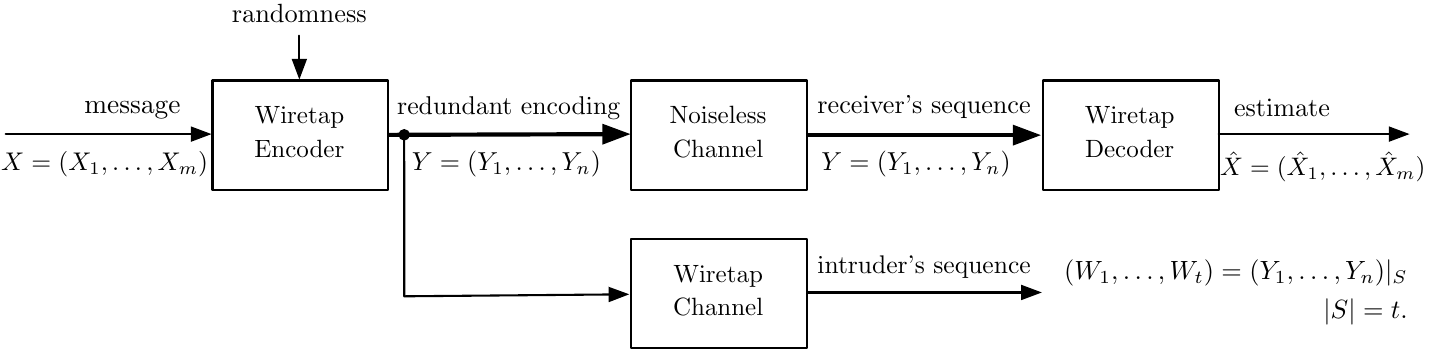}}
\caption[The Wiretap~II Problem]{The Wiretap~II Problem.}
\label{fig:wiretapII}
\end{figure*}

The system defined above is illustrated in Fig.~\ref{fig:wiretapII}.  
The security of the system is defined by the following conditional
entropy, known as ``equivocation'':\index{equivocation} \[ \Delta :=
\min_{S\colon |S| = t}
H(X|W).\] 
When $\Delta = H(X) = m$, the intruder obtains no information about the
transmitted message and we have \emph{perfect privacy} in the
system. Moreover, when $\Delta \to m$ as $m \to \infty$, we call the
system \emph{asymptotically perfectly} private.\index{perfect
  privacy}\index{asymptotically perfect privacy} These two cases
correspond to what is known in the literature as ``strong
secrecy''\index{strong secrecy}\index{weak secrecy}.  A weaker
requirement (known as ``weak secrecy'') would be to have $m - \Delta =
o(m)$.

\begin{rem} \label{rem:randomMessage}
  The assumption that $X$ is sampled from a uniformly random
  source should not be confused with the fact that Alice is
  transmitting \emph{one particular} message to Bob that is fixed and
  known to her before the transmission. In this case, the randomness
  of $X$ in the model captures the \emph{a priori} uncertainty about
  $X$ for the \emph{outside world}, and in particular the intruder,
  but not the transmitter.

  As an intuitive example, suppose that a random key is agreed upon
  between Alice and a trusted third party, and now Alice wishes to
  securely send her particular key to Bob over a wiretapped
  channel. Or, assume that Alice wishes to send an audio stream to Bob
  that is encoded and compressed using a conventional audio encoding
  method.
%

  Furthermore, the particular choice of the distribution on $X$ as a
  uniformly random sequence will cause no loss of generality. If the
  distribution of $X$ is publicly known to be non-uniform, the
  transmitter can use a suitable source-coding scheme to compress the
  source to its entropy prior to the transmission, and ensure that
  from the intruder's point of view, $X$ is uniformly distributed.
  On the other hand, it is also easy to see that if a protocol
  achieves perfect privacy under uniform message distribution, it
  achieves perfect privacy under any other distribution as well.
\end{rem}

The main focus of this paper is on asymptotic trade-offs between the
\emph{rate} $R := m/n$ and the \emph{resilience} $\delta := t/n$ of an asymptotically perfectly
private wiretap coding scheme.  
We will focus on the case where the fraction $\delta$ of the
symbols observed by the intruder is an arbitrary constant below $1$,
which is the most interesting range in our context. However, some of
our constructions work for sub-constant $1-\delta$ as well.

Following \cite{ref:Wyner2}, it is easy to see that, for resilience
$\delta$, an infor\-mation-theoretic bound $R\le 1-\delta + o(1)$ must
hold.  Lower bounds for $R$ in terms of $\delta$ have been studied by
a number of researchers.

For the case of perfect privacy (where the equivocation $\Delta$ is exactly equal to 
the message length), Ozarow and
\index{Ozarow-Wyner's protocol} Wyner~\cite{ref:Wyner2} give a
wiretap coding scheme using linear error-correcting
codes, and show that the existence of an $[n,k,d]_q$-code implies the
existence of a perfectly private wiretap coding scheme with
message length $k$ and block length $n$ (thus, rate $k/n$)
and resilience $\delta = (d-1)/n$.

As a result, the so-called Gilbert-Varshamov bound on the
rate-distance trade-offs of linear codes \cite{gilbert,varshamov}
implies that, asymptotically, $R\ge 1-h_q(\delta)$, where $h_q$ is the
$q$-ary entropy function defined as
\[
h_q(x) := x \log_q(q-1) - x \log_q(x) - (1-x) \log_q(1-x).
\]
If $q\ge 49$ is a square, the bound can be further improved to $R\ge
1-\delta-1/(\sqrt{q}-1)$ using Goppa's algebraic-geometric
codes~\cites{gopp:81,tsvz:82}.  

Moving away from perfect to asymptotically perfect privacy, it was (implicitly)
shown in~\cite{ref:KJS} that for any $\gamma>0$ there exist binary
asymptotically perfectly private wiretap coding schemes with $R\ge
1-2\delta-\gamma$ and exponentially small error.
  This bound strictly improves the coding-theoretic bound of
Ozarow and Wyner for the binary alphabet. 


\subsection{Overview of our Results}

In this paper we prove several lower bounds for the rate of
  asymptotically perfectly private wiretap protocols with
  negligible, i.e., superpolynomially small, error.
  This is shown constructively; i.e., by demonstrating polynomial-time
  computable encoding and decoding schemes that achieve the
  lower bounds.
Our main tool is the design of various types of invertible
  extractors, a concept defined in Section~\ref{sec:InvExt}.
  
  The formal model that our results are based on is defined in Section~\ref{sec:model}.
  As we will see, this model is more stringent than the original Wiretap~II problem.
  In Section~\ref{sec:symb-fix}, we review a known construction of symbol-fixing
  extractors based on linear codes. Then we extend this result to the case of
  ``restricted'' affine sources, using \emph{rank-metric} codes. Using these extractors,
  we are able to recover Ozarow and Wyner's coding-based wiretap schemes as well as a recent
  construction of wiretap schemes for network coding (due to Silva and Kschischang
\cite{ref:RankMetric}, discussed in Section~\ref{sec:NetCod}) in our framework.
    
Our next construction, described in Section~\ref{sec:walk}, shows that if the
  alphabet size is $d$, then there exists
  $\alpha_d\in(0,1)$ such that for every $\eta>0$ and every constant
resilience $\delta \in [0, 1)$, we essentially have rate $R\ge
  \max\{\alpha_d(1-\delta), 1-\delta/\alpha_d\}-\eta$ with
  exponentially small error. 
This is achieved by suitably modifying the symbol-fixing extractor of
  Kamp and Zuckerman~\cite{ref:KZ}.
Contrary to the coding theoretic construction of Ozarow and Wyner,
for a \emph{fixed} alphabet size our bound gives a positive rate for every constant resilience
$\delta \in [0, 1)$.

\begin{figure*}[t]
  \centering
  \begin{tikzpicture}[xscale=4, yscale=4]
    \foreach \x in {1, ..., 9} { \draw[thin, dotted] (0.1*\x,
      1-0.1*\x) -- (0.1*\x, 0); \draw[thin, dotted] (0, 0.1*\x) -- (1
      - 0.1*\x, 0.1*\x); } \draw[->] (-0.02,0) -- (1.1,0) node[right]
    {$\delta$}; \draw[->] (0,-0.02) -- (0,1.1) node[above]
    {$\mathrm{rate}$}; \draw(0, 1) node[left, scale=1]{${1}$};
    \draw(1, 0) node[below, scale=1]{$1$};
    \draw[very thick] (0,1) -- (1,0); 
    \draw[very thick, densely dotted] (0,0.5) -- (1,0); 
    \draw (0,1) -- (0.5,0) node[below]{$\frac{1}{2}$}; \draw[dashed]
    plot[raw gnuplot, id=GV] function{ h(x) = (- x * log(x) - (1-x) *
      log(1-x) ) / log(2); gv(x) = (x == 0) ? 1 : ( (x==0.5) ? 0: 1 -
      h(x)); plot [x=0:0.5] gv(x); };

    \draw[very thick] (0.7, 1.05) node[left]{{\small $(1)$}} -- (0.9,
    1.05); \draw (0.7, 0.9) node[left]{{\small $(2)$}} -- (0.9, 0.9);
    \draw[dashed](0.7, 0.75) node[left]{{\small $(3)$}} -- (0.9,
    0.75); \draw[very thick, densely dotted](0.7, 0.6)
    node[left]{{\small $(4)$}} -- (0.9, 0.6);
  \end{tikzpicture} \hspace{1cm}
  \begin{tikzpicture}[xscale=4, yscale=4]
    \foreach \x in {1, ..., 9} { \draw[thin, dotted] (0.1*\x,
      1-0.1*\x) -- (0.1*\x, 0); \draw[thin, dotted] (0, 0.1*\x) -- (1
      - 0.1*\x, 0.1*\x); } \draw (0.5, 0) node[below]{$\frac{1}{2}$};
    \draw[->] (-0.02,0) -- (1.1,0) node[right] {$\delta$}; \draw[->]
    (0,-0.02) -- (0,1.1) node[above] {$\mathrm{rate}$}; \draw(0, 1)
    node[left, scale=1]{${1}$}; \draw(1, 0) node[below, scale=1]{$1$};
    \draw[very thick] (0,1) -- (1,0); 
    \draw plot[raw gnuplot, id=plot1] function{ d =
      64; 
      d2 = 62; 
      lambda = (2 * sqrt(d2 - 1) + (d - d2)) / d; alpha = -log(lambda
      ** 2)/log(d); max(a, b) = (a > b) ? a : b; walk(x) = max( alpha
      * (1-x), 1 - x / alpha ); plot [x=0:1] walk(x); }; \draw[dashed]
    plot[raw gnuplot, id=plot2] function{ d = 64; 
      h(q, x) = (x * log(q-1) - x * log(x) - (1-x) * log(1-x) ) /
      log(q); max(a, b) = (a > b) ? a : b; gv(x) = (x == 0) ? 1 : (
      (x==1) ? 0: 1 - h(d, x)); AG(x) = 1 - x - (1 / (sqrt(d) - 1));
      plot [x=0:1] max(gv(x), AG(x)); };
    \draw[very thick] (0.7, 1.05) node[left]{{\small $(1)$}} -- (0.9,
    1.05); \draw (0.7, 0.9) node[left]{{\small $(5)$}} -- (0.9, 0.9);
    \draw[dashed](0.7, 0.75) node[left]{{\small $(6)$}} -- (0.9,
    0.75);
  \end{tikzpicture}
  \caption[Comparison of the rate vs.\ resilience trade-offs achieved
  by various wiretap protocols]{ A comparison of the rate vs.\
    resilience trade-offs achieved by the wiretap protocols for the
    binary alphabet (left) and larger alphabets (right, in this
    example of size $64$).  $(1)$ Infor\-mation-theoretic bound,
    attained by Theorem~\ref{coro:wiretap}; $(2)$ The bound approached
    by \cite{ref:KJS}; $(3)$ Protocol based on best non-explicit binary
    linear codes \cites{gilbert,varshamov}; $(4)$ The construction of
    \cite{ref:CDH} (based on all-or-nothing transforms), assuming that 
    the underlying exposure-resilient function is optimal (see Appendix~\ref{app:ERC}
    for a discussion); $(5)$
    Random walk protocol of Corollary~\ref{coro:WTWalk}, using a Ramanujan graph; $(6)$
    Protocol based on the best known explicit \cite{tsvz:82} and
    non-explicit \cites{gilbert,varshamov} linear codes.  }

  \label{fig:region}
\end{figure*}
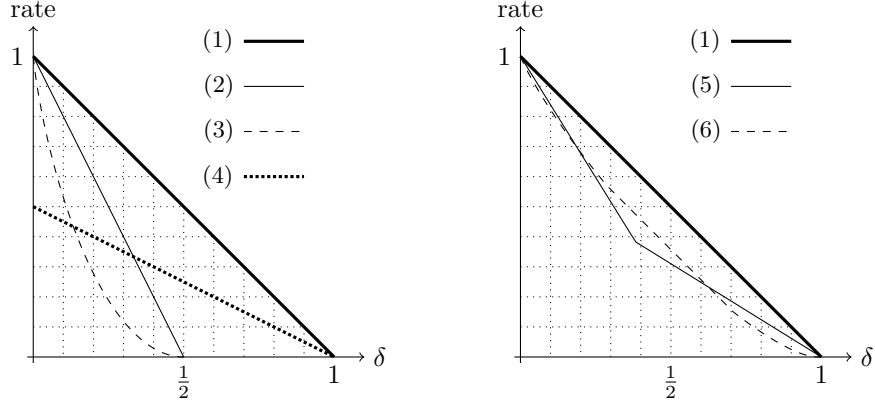

Even though the bound in Section~\ref{sec:walk} is superceded by our
  main result in Section~\ref{sec:invAExt}, we have included it
  because of its simplicity and potential for practical use\footnote{
The construction in Section~\ref{sec:walk} has other
features that are not offered by the result in Section~\ref{sec:invAExt}.
For example, this construction achieves an exponentially small error
(see Definition~\ref{def:wiretap}) and can handle any alphabet size
larger than $2$. In contract, our explicit construction in 
Section~\ref{sec:invAExt} offers super-polynomially small error and
can only be defined for alphabet sizes that are prime powers.
  }.
Our second bound (Theorem~\ref{coro:wiretap}) matches the
  information-theoretic 
  upper bound of Ozarow and Wyner. Namely,
  for any prime power alphabet
  size $q$, and any resilience $\delta\in[0,1)$,
  we construct a wiretap protocol with superpolynomially small 
  error, zero leakage and rate $\ge 1-\delta-o(1)$.
In fact, this bound holds in a more general setting in which the
  intruder is not only allowed to look at a $\delta$-fraction of
  the symbols of Alice's message, but is also allowed to perform
  any linear preprocessing of Alice's message before doing so. 
The power of this result stems largely from a black box transformation
  which makes certain seedless extractors invertible. 
More specifically, the results of this section are obtained by
  applying this transformation to certain affine extractors. 
A plot of the bounds (which also compares our bounds with those
obtained in other relevant works) can be found in Fig.~\ref{fig:region}. 

In sections~\ref{sec:activeIntruder} and \ref{sec:NetCod}, 
we will demonstrate several important applications of
this fact in the context of network coding as well as wiretapped
communication in the presence of noise and active intruders.
In particular we provide, for the
  first time, an optimal solution to the wiretap problem in network
  coding \cite{ref:NetWT} without imposing any restrictions
  (such as a large alphabet or packet length, or any change in the
  network code).  We remark that our coding scheme adds a privacy layer to any
  existing network coding scheme as an outer code, without affecting the existing
  code being used. Thus, the resulting network coding scheme may be
  made resilient against noise (or active intruders) if the original network code can
  handle errors.

  The final application in Section~\ref{subsec:arbitrary} studies an
  all-powerful intruder who is only limited by the amount of
  information he can obtain from Alice's encoded message, and not by
  the nature of the observations.  By inverting seeded extractors with
  nearly-optimal output lengths, we will show that if Alice and Bob
  have access to a \emph{side channel} over which
  Alice can publicly send a polylogarithmic number of bits to Bob (that
  can be seen by the eavesdropper), then their
  communication on the main channel can be made secure even if
  the intruder can access the values of any $t$ Boolean functions of
  Alice's encoded message.

  \section{Preliminaries and Basic Facts}
  \label{sec:notation}

\subsection{Notation and Probability Distributions}

For a prime power $q$, we use $\F_q$ to denote the finite field with
  $q$ elements.  We will occasionally use the notation $\F_2$ for the
  set $\{0,1\}$, even if we do not need to use the field
  structure. For a positive integer $n$, define $[n]$ as the set $\{1,
  2, \ldots, n\}$. For a vector $x = (x_1, x_2, \ldots, x_n)$ and a
  subset $S \subseteq [n]$, we denote by $x|_S$ the vector of length
  $|S|$ that is obtained from $x$ by removing all the coordinates
  $x_i$, $i \notin S$.  For an integer $k > 0$, we will use the
  notation $\U_k$ for the uniform distribution on $\F_2^k$.
  More generally, for a finite set $\Omega$, we will use $\U_\Omega$
  for the uniform distribution on $\Omega$.
  For a function $f$, we denote by $f^{-1}(x)$ the set of the
  preimages of $x$, i.e., the set $\{ y\colon f(y) = x\}$.
  We denote the probability measure defined by a distribution $\cX$ by
  $\Pr_\cX$, hence, $\Pr_\cX(x)$ and $\Pr_\cX[S]$ for $x \in \Omega$
  and $S \subseteq \Omega$ denote the probability that $\cX$ assigns
  to an outcome $x$ and an event $S$, respectively. 
  We will use $\cX|S$ to denote the conditional distribution of $\cX$
  restricted to the set (event) $S$, and $X \sim \cX$ to denote that a random variable $X$ is distributed
  according to $\cX$.

\begin{defn}
  The \emph{support} of a distribution is the set of all the elements
  of the sample space to which it assigns nonzero probabilities.  The
  \emph{min-entropy} of a distribution $\cX$ with finite support $S$
  is defined as
  \[
  H_\infty(\cX) \eqdef \min_{x \in S}\{-\lg \Pr_\cX(x)\},
  \]
  where typically $\lg(\cdot)$ is the logarithm function in base $2$.
  However, when $\cX$ is supported on a set of $d$-ary strings, we
  find it more convenient to use the logarithm function in base $d$
  and measure the entropy in $d$-ary symbols instead of bits.  The
  \emph{Shannon} entropy of the distribution, on the other hand, is
  defined as
  \[
  H(\cX) \eqdef \sum_{x \in S} (-\Pr_\cX(x) \lg \Pr_\cX(x)).
  \]

\end{defn}

When a distribution defined on the set of $n$-bit strings has
min-entropy $k$, the quantity $k/n$ defines the \emph{entropy rate}
of the distribution.  Note that the above definition immediately
implies that the min-entropy of a distribution is upper bounded by its
Shannon entropy (which is in fact the ``expectation'' of the
logarithm of the probabilities).  Hence, if the min-entropy of a
distribution is at least $h$, its Shannon-entropy is also at least
$h$. These two measures however coincide for uniform distributions.

\begin{defn}
  The \emph{statistical distance} (or \emph{total variation distance})
  of two distributions $\cX$ and $\cY$ defined on the same finite
  space $S$ is given by
  \[
  \frac{1}{2} \sum_{s \in S} |\Pr_\cX(s) - \Pr_\cY(s)|,
  \]
  and is denoted by $\dist(\cX, \cY)$.  Note that this is half the
  $\ell_1$ distance of the two distributions when regarded as vectors
  of probabilities over $S$.
\end{defn}

It can be shown that the statistical distance of the two distributions
is at most $\eps$ if and only if for every $T \subseteq S$, we have
$|\Pr_\cX[T] - \Pr_\cY[T]| \leq \eps$. When the statistical distance
of $\cX$ and $\cY$ is at most $\eps$, they are said to be \emph{$\eps$-close}
and this is denoted by $\cX \sim_\eps \cY$.

While we defined the above terms for probability distributions, with a slight abuse of notation 
we may use them interchangeably for random variables as well.

\section{The Formal Model}
\label{sec:model}
The model that we consider in this work is motivated by
the original wiretap channel problem but is more stringent in terms of
its security requirements. In particular, instead of using Shannon
entropy as a measure of uncertainty, we will rely on statistical
indistinguishability which is a stronger measure that is more widely
used in cryptography.

\begin{defn} \index{wiretap protocol}
  \label{def:wiretap}
  Let $\Sigma$ be a set of size\footnote{Throughout the paper,
  when the alphabet size is a prime power, we find it more convenient
  to use the symbol $q$ for the alphabet size.} $d$, parameters $m$ and $n$ be positive integers,
  and $\veps, \gamma>0$.  A $(t,\veps,\gamma)_d$-resilient wiretap
  protocol of block length $n$ and message length $m$ is a pair of
  functions $E\colon\Sigma^m\times\zo^r\to\Sigma^n$ (the encoder) and
  $D\colon\Sigma^n\to\Sigma^m$ (the decoder) that are computable in
  time polynomial in $m$, such that
  \begin{enumerate}
  \item[(a)] (Decodability) For all $x\in\Sigma^m$ and all $z\in\zo^r$
    we have $D(E(x,z))=x$,
  \item[(b)] (Resiliency) Let $\rv{X} \sim \U_{\Sigma^m}$, $\rv{Z}\sim
    \U_{r}$, and $\rv{Y}:=E(X,Z)$. For a set $S \subseteq [n]$ and $w
    \in \Sigma^{|S|}$, let
    $\cX_{S,w}$ denote the distribution of $X$ conditioned on the
    event $Y|_S = w$. Define the set of \emph{bad observations} as \[
    B_S := \{ w \in \Sigma^{|S|} \mid \dist( \cX_{S,w}, \U_{\Sigma^m}
    ) > \eps \},\] where $\dist(\cdot, \cdot)$ denotes the statistical
    distance between two distributions.  Then we require that for
    every $S \subseteq [n]$ of size at most $t$, $\Pr[Y|_S \in B_S]
    \leq \gamma$.
  \end{enumerate}

  The \emph{encoding} of a vector $x\in\Sigma^k$ is accomplished by
  choosing a vector $Z\in\zo^r$ uniformly at random, and calculating
  $E(x,Z)$.  The quantities $R:=m/n$, $\eps$, and $\gamma$ are called
  the \emph{rate}, the \emph{error}, and the \emph{leakage}
  \index{wiretap protocol!rate}\index{wiretap
    protocol!error}\index{wiretap protocol!leakage}\index{wiretap
    protocol!resilience} of the protocol, respectively.  Moreover, we
  call $\delta:=t/n$ the \emph{(relative) resilience} of the protocol.
\end{defn}

The decodability condition ensures that the functions $E$ and $D$ are
a \emph{matching} encoder/decoder pair, while the resiliency
conditions ensures that the intruder learns almost nothing about the
message from his observation.

In our definition, the imperfection of the protocol is captured by the
two parameters $\eps$ and $\gamma$.  When $\eps = \gamma = 0$, the
above definition coincides with the original wiretap channel problem
(as defined by Ozarow and Wyner \cite{ref:Wyner2} and described in
the introduction) for the case of perfect privacy.

When $\gamma = 0$, we will have a \emph{worst-case} guarantee, namely,
that the intruder's views of the message before and after his
observation are statistically close, \emph{regardless} of the outcome
of the observation.

The protocol remains interesting even when $\gamma$ is positive but
sufficiently small.  When $\gamma > 0$, a particular observation might
potentially reveal to the intruder a lot of information about the
message.  However, a negligible $\gamma$ will ensure that such a bad
event (or \emph{leakage}) happens only with negligible probability.

All the constructions that we will study in this paper achieve zero
leakage (i.e., $\gamma = 0$), except for the general result in
Section~\ref{subsec:arbitrary} for which a nonzero leakage is
inevitable.

The significance of zero-leakage protocols is that they assure
\emph{adaptive} resiliency in the \emph{weak} sense introduced in
\cite{ref:DSS01} for exposure-resilient functions.
Notice that the resiliency condition in Definition~\ref{def:wiretap}
can be interpreted as follows: Suppose that the intruder fixes the subset $S$
of the positions to be observed
\emph{before} the protocol runs. Then, when the encoded string $Y$ is created
from the message, the intruder learns the subsequence $W := Y|_S$. Now, suppose that 
at this point a third party called ``the challenger'' randomly presents the intruder with a string
that is either a uniformly random string or the original message $X$
(where the two cases are equally likely). The resiliency condition
of Definition~\ref{def:wiretap} essentially ensures that the intruder will not
be able to distinguish between the two cases better than a random guess
(simply because the distribution of $X$ conditioned on the observation
is statistically close to uniform). However, in general, some ``bad''
outcomes of $W$ may reveal non-negligible information about $X$ to the intruder
(by definition, this happens with a small probability $\gamma$).
 Now consider an \emph{adaptive} intruder who does not fix $S$ beforehand,
 but observes $W$ adaptively after the encoding $Y$ is created.
 This means that the choice of each query made by the intruder may depend
on the outcome of the previous queries (i.e., those positions in $W$ that are already
revealed). Now in this adaptive setting, the intruder has enough power
to potentially \emph{direct} the observation towards bad outcomes, by choosing
the query positions smartly. However, when the leakage parameter $\gamma$ is zero,
we can ensure that this cannot happen and the intruder will not be able to respond the
challenge much better than a random guess even when allowed to make adaptive queries.

In general, it is straightforward to verify that our model can be used
to solve the original wiretap~II problem, with $\Delta \geq
m(1-\eps-\gamma)$:

\begin{lem}
  \label{app:model}
  Suppose that $(E, D)$ is an encoder/decoder pair as in
  Definition~\ref{def:wiretap}.  Then using $E$ and $D$ in the
  wiretap~II problem attains an equivocation \[ \Delta \geq
  m(1-\eps-\gamma).\]
\end{lem}

\begin{proof}
  Let $W := Y|_S$ be the intruder's observation, and denote by $W'$
  the set of \emph{good} observations, namely,
  \[ W' := \{ w \in \Sigma^t \colon \dist( \cX_{S,w}, \U_{\Sigma^m} )
  \leq \eps \}. \] Denote by $H(\cdot)$ the Shannon entropy in $d$-ary
  symbols. Then we will have
  \begin{eqnarray*}
    H(X|W) &=& \sum_{w \in \Sigma^t} \Pr(W = w) H(X|W = w) \\
    &\geq& \sum_{w \in W'} \Pr(W = w) H(X|W = w) \\
    &\stackrel{\mathrm{(a)}}{\geq}& \sum_{w \in W'} \Pr(W = w) (1-\eps) m  \\
    &\stackrel{\mathrm{(b)}}{\geq}& (1-\gamma)(1-\eps) m \geq (1-\gamma-\eps) m. 
  \end{eqnarray*}
  The inequality $\mathrm{(a)}$ follows from the definition of $W'$
  combined with Proposition~\ref{prop:Shannon} in the appendix, and
  $\mathrm{(b)}$ by the definition of the leakage parameter.
\end{proof}

Hence, we will achieve asymptotically perfect privacy when
$\eps+\gamma = o(1/m)$.  For all the protocols that we present in this
work this quantity will be superpolynomially small; that is,
smaller than $1/m^c$ for every positive constant $c$ (provided that
$m$ is large enough).


There are several interrelated notions in the literature on
Cryptography and Theoretical Computer Science that are also closely
related to our definition of the wiretap protocol
(Definition~\ref{def:wiretap}).  These are \emph{resilient functions
  (RF)} and \emph{almost perfect resilient functions (APRF)},
\emph{exposure-resilient functions (ERF)}, and \emph{all-or-nothing
  transforms (AONT)} (cf.\
\cites{ref:tResilient,ref:tResilient2,ref:Riv97,ref:Stinson,ref:FT00,ref:CDH,ref:KJS}
and \cite{ref:Dodis} for a comprehensive account of several important
results in this area). We have included a short survey of these notions
that includes a comparisons with our model in Appendix~\ref{app:ERC}.

Among these, the notion of AONTs is the most relevant to our work. Roughly speaking,
an AONT is an efficiently invertible randomized function mapping
$\zo^m$ to $\zo^n$ ($n \geq m$) such that
the joint distribution of the output bits on any set of up to $t$
coordinate positions is (nearly) independent of the input.
As discussed in Appendix~\ref{app:ERC}, AONTs can be used in the
original wiretap~II model as the encoder/decoder pair. However,
this notion turns out to be stronger than our notion of wiretap protocols
and the best known explicit constructions of AONTs achieve a substantially
sub-optimal rate/resilience tradeoff (see Fig.~\ref{fig:region}).

\section{Randomness Extractors and Constructions}
\label{sec:symb-fix}

A combinatorial tool that is of central importance in our constructions
of wiretap protocols is the notion of \emph{randomness extractors}.
In this section, we first review some standard definitions and facts in the theory
of randomness extractors that are relevant to our work (Section~\ref{sec:extr-prelim}).
We refer the reader to \cite{ref:survey} for a more detailed account of these notions.
Then, in Section~\ref{sec:zeroerror-extr}
 we present some explicit constructions that will be used for explicit
construction of perfectly private wiretap protocols.

\subsection{Preliminaries on Extractors}
\label{sec:extr-prelim}

Before we present a formal definition of randomness extractors,
we define \emph{families} of random sources, as follows.

\begin{defn}
  Let $\Sigma$ be a finite set of size $d > 1$.  An $(n, k)_d$
  \emph{family} of $d$-ary randomness sources of length $n$ and
  min-entropy $k$ is a set $\mathcal{F}$ of probability distributions
  on $\Sigma^n$ such that every $\cX \in \mathcal{F}$ has $d$-ary
  min-entropy at least $k$ .
\end{defn}

There are numerous natural families of sources that have been
introduced and studied in the theory of randomness extractors.  In
this work, besides the general family of distributions with high
min-entropy, we will focus on the family of \emph{symbol-fixing} and
\emph{affine} sources, defined below.

\begin{defn}
  An $(n, k)_d$ \emph{symbol-fixing} source is the distribution of a
  random variable $$\rv{X}=(\rv{X}_1, \rv{X}_2, \ldots, \rv{X}_n) \in
  \Sigma^n,$$ for some set $\Sigma$ of size $d$, in which at least $k$
  of the coordinates (chosen arbitrarily) are uniformly and
  independently distributed on $\Sigma$ and the rest take
  deterministic values.

  When $d=2$, we will have a \emph{binary} symbol-fixing source, or
  simply a \emph{bit-fixing} source. In this case $\Sigma = \{0, 1\}$,
  and the subscript $d$ is dropped from the notation.
\end{defn}

\begin{defn}
  For a prime power $q$, the family of $q$-ary $k$-dimensional
  \emph{affine sources} of length $n$ is the set of distributions on
  $\F_q^n$, each uniformly distributed on an affine translation of
  some $k$-dimensional sub-space of $\F_q^n$.
\end{defn}

Affine sources are natural generalizations of symbol-fixing sources
when the alphabet size is a prime power. It is easy to see that the
$q$-ary min-entropy of a $k$-dimensional affine source is $k$.

\newcommand{\extf}{f} 

\begin{defn}
  A function $\extf\colon \zo^n \times \zo^r \to \zo^m$ is a strong
  seeded $(k, \eps)$-extractor if for every distribution $\cX$ on
  $\zo^n$ with min-entropy at least $k$, random variable $X \sim \cX$
  and a \emph{seed} $Z \sim \U_r$, the distribution of $(\extf(X, Z), Z)$
  is $\eps$-close to $\U_{m+r}$. An extractor is \emph{explicit} if it
  is polynomial-time computable.
\end{defn}

A strong extractor $\extf(x, z)$ for a source $\cX$ with error $\eps^2$
satisfies the property that for all but a $\eps$ fraction of the choices
of the seed $z$, the distribution of $\extf(\cX, z)$ is $\eps$-close to
uniform. This is easily seen by an averaging argument.

For more restricted sources (in particular, symbol-fixing and 
affine sources), \emph{seedless} (or
\emph{deterministic}) extraction is possible. 

\begin{defn}
  Let $\Sigma$ be a finite alphabet of size $d > 1$.  A function
  $\extf\colon \Sigma^n \to \Sigma^m$ is a (seedless) $(k,
  \eps)_d$-extractor for a family $\mathcal{F}$ of $(n, k)_d$ sources
  (defined on $\Sigma^n$) if for every distribution $\cX \in
  \mathcal{F}$ with $d$-ary min-entropy at least $k$, the distribution
  $\extf(\cX)$ is $\eps$-close to $\U_{\Sigma^m}$. A seedless extractor is
  \emph{explicit} if it is polynomial-time constructible.
\end{defn}

Over large fields, the following affine extractor due to Gabizon and
Raz extract almost the entire source entropy:

\newcommand{\GR}{\mathsf{GR}} 

\begin{thm}\cite{ref:affine}
  There is a constant $q_0$ such that for any prime power field
  size $q$ and integers $n,k$ such that $q > \max\{q_0, n^{20}\}$,
  there is an explicit affine $(k, \eps)_q$-extractor $\GR\colon \F_q^n
  \to \F_q^{k-1}$, where $\eps < q^{-1/21}$. \qed
\end{thm}

In this construction, the field size has to be polynomially large in
$n$. When the field size is small (in particular, constant), the task
becomes much more challenging. The most challenging case thus
corresponds to the binary field $\F_2$, for which an explicit affine
extractor was obtained, when the input entropy is a constant fraction
of the input length, by Bourgain:

\newcommand{\aext}{\mathsf{AExt}} \newcommand{\ibou}{\mathsf{AExt}}

\begin{thm} \cite{ref:Bourgain} \label{thm:Bourgain} For every
  constant $0 < \delta \leq 1$, there is an explicit affine extractor
  $\bou\colon \F_2^n \to \F_2^m$ for min-entropy $\delta n$ with
  output length $m = \Omega(n)$ and error at most
  $2^{-\Omega(m)}$. \qed
\end{thm}
Bourgain's construction was later simplified, improved, and
extended to work for arbitrary prime fields by Yehudayoff
\cite{ref:Yeh09}. We remark that, for the case of binary field,
one case also use a more recent construction of affine extractors
due to Li~\cite{ref:Li11} that obtains comparable parameters to
Theorem~\ref{thm:Bourgain} using simpler techniques.


\subsection{Symbol-Fixing and Affine Extractors from Codes}
\label{sec:zeroerror-extr}

We now see simple constructions of zero-error, symbol-fixing
and affine extractors using linear functions arising from good
error-correcting codes. These extractors achieve the lowest possible
error, but however are unable to extract the entire source
entropy over small alphabets. Moreover, the affine extractor only works for a
``restricted'' class of affine sources. 
However, these extractors have the nice property of being
linear, a property that turns out to be useful for our construction of
explicit wiretap schemes discussed in the following sections.

\subsubsection{Symbol-Fixing Extractors from Linear
  Codes} \label{sec:symfix-code}

The theorem below (proved, with a different terminology, in \cite{ref:tResilient}) states that linear error-correcting codes can
be used to obtain symbol-fixing extractors with zero error.

\begin{thm} \label{thm:symfixExtr} Let $\C$ be an $[n,\Tk,d]_q$ code
  over $\F_q$ and $G$ be a $\Tk \times n$ generator matrix of $\C$.
  Then, the function $E\colon \F_q^n \to \F_q^\Tk$ defined
  as\footnote{We typically consider vectors be represented in row
    form, and use the transpose operator
    ($x^\top$)\index{notation!$x^\top$} to represent column vectors.}
  $E(x) := G x^\top$ is an $(n-d+1, 0)_q$-extractor for symbol-fixing
  sources over $\F_q$.

  Conversely, if a linear function $E\colon \F_q^n \to \F_q^\Tk$ is an
  $(n-d+1, 0)_q$-extractor for symbol-fixing sources over $\F_q$, it
  corresponds to a generator matrix of an $[n,\Tk,d]_q$ code.
\end{thm}

\begin{proof}
  Let $\cX$ be a symbol-fixing source with a set $S \subseteq [n]$ of
  fixed coordinates, where\footnote{If the set of fixed symbols if of
    size smaller than $d-1$, the argument still goes through by taking
    $S$ as an arbitrary set of size $d-1$ containing all the fixed
    coordinates.}  $|S| = d-1$, and define $\bar{S} := [n] \setminus
  S$.  Observe that, by the Singleton bound, we must have $|\bar{S}| =
  n-d+1 \geq \Tk$.
 
  The submatrix of $G$ obtained by removing the columns picked by $S$
  must have rank $\Tk$. Since otherwise, the left kernel of this
  submatrix would be nonzero, meaning that $\C$ has a nonzero codeword
  that consists of entirely zeros at the $d-1$ positions picked by
  $S$, contradicting the assumption that the minimum distance of $\C$
  is $d$.  Therefore, the distribution $E(\cX)$ is supported on a
  $\Tk$-dimensional affine space on $\F_q^\Tk$, meaning that this
  distribution is uniform.
 
  The converse is straightforward by following the same argument.
\end{proof}

If the field size is large enough; e.g., $q \geq n$, then one can pick
$\C$ in the above theorem to be an MDS code (in particular, a
Reed-Solomon code) to obtain a $(k, 0)$-extractor for all
symbol-fixing sources of entropy $k$ with optimal output length
$k$. However, for a fixed $q$, negative results on the rate-distance
trade-offs of codes (e.g., Hamming and Plotkin bounds, as well as the linear programming
bound due to McEliece-Rodemich-Rumsey-Welch, cf.~\cite[Chapter~5]{vanlint}) assert
that this construction of extractors must inevitably lose some
fraction of the entropy of the source. Moreover, the construction
would at best be able to extract some constant fraction of the source
entropy only if the entropy of the source (in $q$-ary symbols) is
above $n/q$.

\subsubsection{Restricted Affine Extractors from Rank-Metric
  Codes} \label{sec:affext-code}

In Section~\ref{sec:Apps}, we will see that affine extractors
can be used to construct wiretap schemes for models that are
more general than the original Wiretap~II problem, e.g., when
the direct channel is noisy. For these applications, the extractor
needs to additionally have a nice structure that is in particular
offered by linear functions. 

An obvious observation is that a nontrivial affine extractor cannot be
a linear function.  Indeed, a linear function $f(x) := \langle
\alpha, x \rangle + \beta$, where $\alpha, \beta, x \in \F_q^n$, is
constant on the $(n-1)$-dimensional orthogonal subspace of $\alpha$,
and thus, fails to be an extractor for even $(n-1)$-dimensional affine
spaces. However, in this section we will see that linear affine extractors
can be constructed if the affine source is known to be described by
a set of linear constraints whose coefficients lie on a small \emph{sub-field} 
of the underlying field.
Such restricted extractors turn out to be sufficient for some of the
applications that we will consider.

Let $Q$ be a prime power.  Same as linear codes, an affine subspace on
$\F_Q^n$ can be represented by a \emph{generator matrix}, or
\emph{parity-check matrix} and a constant shift. That is, a
$k$-dimensional affine subspace $A \subseteq \F_Q^n$ can be described
as the image of a linear mapping
\[
A := \{ x G + \beta \colon x \in \F_Q^k \},
\]
where $G$ is a $k \times n$ \emph{generator matrix} of rank $k$ over
$\F_Q$, and $\beta \in \F_Q^n$ is a fixed vector. Alternatively, $A$
can be expressed as the translated null-space of a linear mapping
\[
A := \{ x + \beta \in \F_Q^n\colon H x^\top = 0 \},
\]
for an $(n-k) \times n$ \emph{parity check matrix} of rank $n-k$ over
$\F_Q$.

Observe that a symbol-fixing source over $\F_q$ with $q$-ary
min-entropy $k$ can be seen as a $k$-dimensional affine source with a
generator matrix of the form $[I \mid \mathbf{0}]\cdot P$, where $I$
is the $k \times k$ identity matrix, $\mathbf{0}$ denotes the $k
\times (n-k)$ all-zeros matrix, and $P$ is a permutation matrix.
Recall that from Theorem~\ref{thm:symfixExtr} we know that for this
restricted type of affine sources linear extractors exist. In this
section we generalize this idea.

Suppose that $Q = q^m$ for a prime power $q$ so that $\F_Q$ can be
regarded as a degree $m$ extension of $\F_q$ (and isomorphic to
$\F_{q^m}$).  Let $A$ be an affine source over $\F_Q^n$. We will call
the affine source
\emph{$\F_q$-restricted}\index{source!affine!$\F_q$-restricted} if its
support can be represented by a generator matrix (or equivalently, a
parity check matrix) over $\F_q$.

In this section we introduce an affine extractor that is $\F_Q$-linear
and, assuming that $m$ is sufficiently large, extracts from
$\F_q$-restricted affine sources.  The construction of the extractor
is similar to Theorem~\ref{thm:symfixExtr}, except that instead of an
error-correcting code defined over the \emph{Hamming metric}, we will
use \emph{rank-metric} codes.

Consider the function $\rdist\colon \F_q^{m \times n} \times \F_q^{m
  \times n} \to \Z$, where $\F_q^{m \times n}$ denotes the set of $m
\times n$ matrices over $\F_q$, defined as $\rdist(A, B) :=
\rk_q(A-B)$, where $\rk_q$ is the matrix rank over $\F_q$.  It is
straightforward to see that $\rdist$ is a metric\index{rank metric}.

The usual notion of error-correcting codes defined under the Hamming
metric can be naturally extended to the rank metric.  In particular, a
\emph{rank-metric}\index{code!rank-metric} code $\C$ can be defined as
a set of $m \times n$ matrices (known as codewords), whose minimum
distance is the minimum rank distance between pairs of codewords.

For $Q := q^m$, there is a natural correspondence between $m\times n$
matrices over $\F_q$ and vectors of length $n$ over $\F_Q$. Consider
an isomorphism $\varphi\colon \F_Q \to \F_q^m$ between $\F_Q$ and
$\F_q^m$ which maps elements of $\F_Q$ to column vectors of length $m$
over $\F_q$. Then one can define a mapping $\Phi\colon \F_Q^n \to
\F_q^{m \times n}$ defined as
\[
\Phi(x_1, \ldots, x_n) := [\varphi(x_1) \mid \cdots \mid \varphi(x_n)]
\]
to put the elements of $\F_Q^n$ in one-to-one correspondence with $m
\times n$ matrices over $\F_q$.

A particular class of rank-metric codes are linear ones. Suppose that
$\C$ is a linear $[n,\Tk,\tilde{d}]_Q$ code over $\F_Q$. Then, using $\Phi(\cdot)$, $\C$ can
be regarded as a rank-metric code of dimension $\Tk$ over $\F_q^{m
  \times n}$. In symbols, we will denote such a linear $\Tk$-dimensional
rank-metric code as an $[[ n,\Tk,d ]]_{q^m}$ code, where $d$ is the
minimum rank-distance of the code. The rank-distance of a linear
rank-metric code turns out to be equal to the minimum rank of its
nonzero codewords and obviously, one must have $d \leq \tilde{d}$. However,
the Hamming distance of $\C$ might turn out to be much larger than its
rank distance when regarded as a rank-metric code. In particular, $d
\leq m$, and thus, $d$ must be strictly smaller than $\tilde{d}$ when the
degree $m$ of the field extension is less than $\tilde{d}$.

A counterpart of the Singleton bound in the rank-metric states that,
for any $[[n,\Tk,d]]_{q^m}$ code, one must have $d \leq
n-\Tk+1$. Rank-metric codes that attain equality exist and are called
\emph{maximum rank distance (MRD)}\index{code!MRD} codes. A class of
linear rank-metric codes known as \emph{Gabidulin
  codes}\index{code!Gabidulin} \cite{ref:Gab85} are MRD and can be
thought of as the counterpart of Reed-Solomon codes in the rank
metric. In particular, the codewords of a Gabidulin code, seen as
vectors over the extension field, are evaluation vectors of
bounded-degree \emph{linearized} polynomials rather than arbitrary
polynomials as in the case of Reed-Solomon codes.  These codes are
defined for any choice of $n, \Tk, q, m$ as long as $m \geq n$ and $\Tk
\leq n$.

The following is an extension of Theorem~\ref{thm:symfixExtr} to
restricted affine sources.

\begin{thm} \label{thm:resAffExtr} Let $\C$ be an $[[n,\Tk,d]]_{q^m}$
  code defined from a code over $\F_Q$ (where $Q := q^m$) with a
  generator matrix $G \in \F_Q^{\Tk \times n}$.  Then the function
  $E\colon \F_Q^n \to \F_Q^\Tk$ defined as $E(x) := G x^\top$ is an
  $(n-d+1, 0)$-extractor for $\F_q$-restricted affine sources over
  $\F_Q$.

  Conversely, if a linear function $E\colon \F_Q^n \to \F_Q^\Tk$ is an
  $(n-d+1, 0)$-extractor for all $\F_q$-restricted affine sources over
  $\F_Q$, it corresponds to a generator matrix of an
  $[[n,\Tk,d]]_{q^m}$ code.
\end{thm}

\begin{proof}
  Consider a restricted affine source $\cX$ uniformly supported on an
  affine subspace of dimension\footnote{ The argument still holds if
    the dimension of $\cX$ is more than $n-d+1$.  } $n-d+1$
  \[ X := \{ x A+\beta\colon x \in \F_Q^{n-d+1} \}, \] where $A \in
  \F_q^{(n-d+1) \times n}$ has rank $n-d+1$, and $\beta \in \F_Q^n$ is
  a fixed translation.  Note that $\Tk \leq n-d+1$ by the Singleton
  bound for rank-metric codes.

  The output of the extractor is thus uniformly supported on the
  affine subspace
  \[
  B := \{ G A^\top x^\top + G \beta^\top\colon x \in \F_Q^{n-d+1} \}
  \subseteq \F_Q^\Tk.
  \]

  Note that $G A^\top \in \F_Q^{\Tk \times (n-d+1)}$.  Our goal is to
  show that the dimension of $B$ is equal to $\Tk$.  Suppose not, then
  we must have $\rk_Q (GA^\top) < \Tk$. In particular, there is a
  nonzero $y \in \F_Q^\Tk$ such that $y G A^\top = 0$.

  Let $Y := \Phi(y G) \in \F_q^{m \times n}$, where $\Phi(\cdot)$ is
  the isomorphism that maps codewords of $\C$ to their matrix form
  over $\F_q$. By the distance of $\C$, we know that $\rk_q(Y) \geq
  d$. Since $m \geq d$, this means that $Y$ has at least $d$ linearly
  independent rows. On the other hand, we know that the matrix $Y
  A^\top \in \F_q^{\Tk \times (n-d+1)}$ is the zero matrix. Therefore,
  $Y$ has $d$ independent rows (each in $\F_q^n$) that are all
  orthogonal to the $n-d+1$ independent rows of $A$. Since $d +
  (n-d+1) > n$, this is a contradiction.

  Therefore, the dimension of $B$ is exactly $\Tk$, meaning that the
  output distribution of the extractor is indeed uniform.  The
  converse is straightforward by following a similar line of argument.
\end{proof}

Thus, in particular, we see that generator matrices of MRD codes can
be used to construct linear extractors for restricted affine sources
that extract the entire source entropy with zero error. This is
possible provided that the field size is large enough compared to the
field size required to describe the generator matrix of the affine
source. Using Gabidulin's rank metric codes, we immediately obtain the following
corollary of Theorem~\ref{thm:resAffExtr}:

\begin{coro} \label{coro:resAffExtrGabidulin}
  Let $q$ be a prime power. Then for every positive integer $n$,
  $k \leq n$, and $Q := q^n$, there is a linear function $f\colon \F_Q^n \to \F_Q^k$
  that is an explicit\footnote{We have implicitly assumed that an explicit respresentation
  of the finite field $\F_q$ (i.e., a deterministic polynomial time algorithm for addition,
  multiplication, and encoding of the elements over $\F_q$) is available. This is known to be the case
  for all prime powers of small characteristic \cite{ref:Shoup}.} $(k, 0)$-extractor for $\F_q$-restricted affine sources
  over $\F_Q$. \qed
\end{coro}





\section{Inverting Extractors}
\label{sec:InvExt}

In this section we will introduce the notion of \emph{invertible
  extractors} and its connection with wiretap protocols\footnote{
  Another notion of invertible extractors was introduced in
  \cite{ref:Dod05} and used in \cite{ref:DS05} for a different
  application (entropic security) that should not be confused with the
  one we use.  Their notion applies to seeded extractors with long
  seeds that are efficiently invertible bijections for every fixed
  seed. Such extractors can be seen as a single-step walk on highly
  expanding graphs that mix in one step.  This is in a way similar to
  the multiple-step random walk used in the seedless extractor of
  section \ref{sec:walk}, that can be regarded as a single-step walk
  on the expander graph raised to a certain power.  }.  Later we will
use this connection to construct wiretap protocols with good
rate-resilience trade-offs.

\newcommand{\invp}{\nu} 

\begin{defn} \index{invertible extractor}
  \label{def:inverter}
  Let $\Sigma$ be a finite alphabet and $f$ be a mapping from
  $\Sigma^n$ to $\Sigma^m$.  For $\invp \geq 0$, a function $A\colon
  \Sigma^m \times \zo^r \to \Sigma^n$ is called a
  $\invp$-\emph{inverter} for $f$ if the following conditions hold:
  \begin{enumerate}
  \item[(a)] (Inversion) Given $x \in \Sigma^m$ such that $f^{-1}(x)$
    is nonempty, for every $z \in \zo^r$ we have $f(A(x, z)) =x
    $.
  \item[(b)] (Uniformity) $A(\U_{\Sigma^m}, \U_r) \sim_\invp
    \U_{\Sigma^n}$.
  \end{enumerate}
  A $\invp$-inverter is called \emph{efficient} if there is a
  randomized algorithm that runs in worst case polynomial time and,
  given $x \in \Sigma^m$ and $z$ as a random seed, computes $A(x,
  z)$. We call a mapping $\invp$-\emph{invertible} if it has an
  efficient $\invp$-inverter, and drop the prefix $\invp$ from the
  notation when it is zero.
\end{defn}

The parameter $r$ in the above definition captures the amount of random
bits that the inverter (seen as a randomized algorithm) needs to receive.
For our applications, no particular care is needed to optimize this
parameter and, as long as $r$ is polynomially bounded in $n$, it is generally
ignored (the same remark applies to the parameter $r$ in 
Definition~\ref{def:wiretap}). In this work, we are interested in
randomness extractors that are invertible functions (as in Definition~\ref{def:inverter}).
Such functions will be called ``invertible extractors''.

\begin{rem}
  If a function $f$ maps the uniform distribution to a distribution
  that is $\eps$-close to uniform (as is the case for all extractors),
  then any randomized mapping that maps its input $x$ to a
  distribution that is $\bar{\invp}$-close to the uniform distribution on
  $f^{-1}(x)$, for some $\bar{\invp} \geq 0$, is easily seen to be an $(\eps+\bar{\invp})$-inverter for
  $f$.  In some situations designing such a function might be easier
  than directly following the above definition.
\end{rem}

\begin{rem} \label{rem:linearInvertible}
A \emph{linear} function $f\colon \F_q^n \to \F_q^m$ over a finite field 
$\F_q$ is easily seen to be (perfectly) invertible\footnote{We have
implicitly assumed that the field operations are efficiently computable.}. 
To see this, observe that the inverter is itself
a linear function (of the input and the random seed) that
can be efficiently computed using elementary methods from linear algebra\footnote{
More precisely, suppose that $f(x) := A \cdot x$ where $x \in \F_q^n$ and
$A$ is an $m \times n$ matrix of rank $m$. Add $n-m$ rows to $A$
to obtain an invertible $n \times n$ matrix $\bar{A}$. Then the inverse
function will be given by $\bar{f}(y) := \bar{A}^{-1} \cdot (y, r)$,
where $y \in \F_q^m$ is its input and $r \in \F_q^{n-m}$ is chosen uniformly at random.
}.
In particular, the linear extractors of Theorem~\ref{thm:symfixExtr}
and Theorem~\ref{thm:resAffExtr} are both invertible.
\end{rem}
 
The idea of random pre-image sampling was proposed in \cite{ref:DSS01}
for construction of adaptive AONTs from APRFs. However, they ignored
the efficiency of the inversion, as their goal was to show the
existence of (not necessarily efficient) infor\-mation-theoretically
optimal adaptive AONTs.  Moreover, the strong notion of APRF and a
perfectly uniform sampler is necessary for their construction of
AONTs.  As wiretap protocols are weaker than (worst-case) AONTs, they
can be constructed from slightly imperfect inverters as shown by the
following result.


\begin{thm} \label{thm:protocol} Let $\Sigma$ be an alphabet of size
  $d > 1$ and $\extf\colon \Sigma^n \to \Sigma^m$ be a
  $(\gamma^2/2)$-invertible $d$-ary $(k, \eps)$ symbol-fixing
  extractor.  Then, $\extf$ and its inverter can be seen as a
  decoder/encoder pair for an $(n-k,\veps+\gamma,\gamma)_q$-resilient
  wiretap protocol with block length $n$ and message length $m$. 
\end{thm}

\begin{proof}
  Let $E$ and $D$ denote the wiretap encoder and decoder,
  respectively.  Hence, $E$ is the $(\gamma^2/2)$-inverter for $f$,
  and $D$ is the extractor $f$ itself.  From the definition of the
  inverter, for every $x \in \Sigma^m$ and every random seed $z$, we
  have $D(E(x, z)) = x$.  Hence it is sufficient to show that the pair
  satisfies the resiliency condition.
  
  Let the random variable $X$ be uniformly distributed on $\Sigma^m$
  and the seed $Z\in\zo^r$ be chosen uniformly at random.  Denote the
  encoding of $X$ by $Y := E(X, Z)$. Fix any $S \subseteq [n]$ of size
  at most $n-k$.

  For every $w \in \Sigma^{|S|}$, let $Y_w$ denote the set $\{ y \in
  \Sigma^n\colon (y|_S) = w \}$.
  Note that the sets $Y_w$ partition the space $\Sigma^n$ into
  $|\Sigma|^{|S|}$ disjoint sets.

  Let $\cY$ and $\cY_S$ denote the distribution of $Y$ and $Y|_S$,
  respectively. The inverter guarantees that $\cY$ is
  $(\gamma^2/2)$-close to uniform. Applying
  Proposition~\ref{prop:conditioning} in the appendix, we get that
  \[
  \sum_{w \in \Sigma^{|S|}} \Pr[(Y|_S) = w] \cdot \dist( (\cY | Y_w),
  \U_{Y_w} ) \leq \gamma^2.
  \]
  The left hand side is the expectation of $\dist( (\cY | Y_w),
  \U_{Y_w} )$.  Denote by $W$ the set of all \emph{bad outcomes} of
  $Y|_S$, i.e.,
  \[W := \{ w \in \Sigma^{|S|} \mid \dist( (\cY | Y_w), \U_{Y_w} ) >
  \gamma \}.\] By Markov's inequality, we conclude that
  \[
  \Pr[(Y|_S) \in W] \leq \gamma.
  \]
  For every $w \in W$, the distribution of $Y$ conditioned on the
  event $(Y|_S) = w$ is $\gamma$-close to a symbol-fixing source with
  $n-|S| \geq k$ random symbols. The fact that $D$ is a symbol-fixing
  extractor for this entropy and Proposition~\ref{prop:closeFunction} in the appendix
  imply that, for the fixed choice of $w$, the distribution of
  $D(Y)$ conditioned on the event $(Y|_S) = w$ is $(\gamma+\eps)$-close to uniform.  Hence with
  probability at least $1-\gamma$ the distribution of $X$ conditioned
  on the outcome of $Y|_S$ is $(\gamma+\eps)$-close to uniform. This
  ensures the resiliency of the protocol.
\end{proof}

By combining Theorem~\ref{thm:protocol}, Theorem~\ref{thm:symfixExtr}
using a Reed-Solomon code, and Remark~\ref{rem:linearInvertible}, 
we can obtain a perfectly private,
rate-optimal, wiretap protocol for the Wiretap~II problem over large
alphabets (of size $n$ or larger). This recovers the original result of
Ozarow and Wyner\footnote{In fact, Ozarow and Wyner use a parity check
  matrix of an MDS code in their construction, which is indeed a
  generator matrix for the dual code which is itself
  MDS.}~\cite{ref:Wyner2}:
  
\begin{coro} \label{coro:wiretapii} For every positive integer $n$,
  prime power $q \geq n$, and $\delta \in [0, 1)$, 
  there is a $(\delta n, 0, 0)_q$-resilient wiretap protocol with block length $n$ and rate $1 -
  \delta$ that attains perfect privacy. \qed
\end{coro}

\section{A Wiretap Protocol Based on Random Walks} \label{sec:walk}

In this section we describe a wiretap protocol that achieves a rate
$R$ within a constant fraction of the information theoretically
optimal value $1-\delta$ (the constant depending on the alphabet
size).

To achieve our result, we will modify the symbol-fixing extractor of
Kamp and Zuckerman \cite{ref:KZ}, that is based on random walks on
expander graphs, to make it efficiently invertible without affecting
its extraction properties, and then apply Theorem~\ref{thm:protocol}
above to obtain the desired wiretap protocol.
In the following sub-sections, we first review the preliminaries
on expander graphs that we will need, and then introduce the construction
and its analysis.

\subsection{Preliminaries on Expander Graphs}
\label{sec:preExpander}
For the wiretap protocol constructed in Section~\ref{sec:walk} we need
essentially the same tools used for the symbol-fixing extractor
construction of \cite{ref:KZ}, that we briefly review here.
For a detailed review of the theory of
expander graphs, refer to the excellent survey by Hoory, Linial and
Wigderson \cite{HLW06}, and books \cites{ref:MR,ref:MU}.

We will be working with directed regular expander graphs that are
obtained from undirected graphs by replacing each undirected edge with
two directed edges in opposite directions.  Let $G=(V, E)$ be a
$d$-regular graph.  Then a \emph{labeling} of the edges of $G$ is a
function $L\colon V \times [d] \to V$ such that for every $u \in V$
and $t \in [d]$, the edge $(u, L(u, t))$ is in $E$.  The labeling is
\emph{consistent}\index{consistent labeling} if whenever $L(u, t) =
L(v, t)$, then $u = v$.  Note that the natural labeling of a Cayley
graph (cf.\ \cite{HLW06}) is in fact consistent.

A \emph{family} of $d$-regular graphs 
is an infinite set of $d$-regular graphs such that for every $N \in \N$, the
set contains a graph with at least $N$ vertices.  For a parameter $c
\geq 1$, we will call a family $c$-dense if there is an $N_0 \in \N$
such that, for every $N \geq N_0$, the family has a graph with at
least $N$ and at most $cN$ vertices.  We call a family of graphs
\emph{constructible} if all the graphs in the family have a consistent
labeling that is efficiently computable. That is, there is a uniform,
polynomial-time algorithm that, given $N \in \N$ and $i \in [N], j \in
[d]$, outputs the label of the $j$th neighbor of the $i$th vertex,
under a consistent labeling, in the graph in the family that has $N$
vertices (provided that it exists).

Let $A$ denote the normalized adjacency matrix of a $d$-regular graph
$G$ (that is, the adjacency matrix with all the entries divided by $d$).
We denote by $\lambda_G$ the second largest eigenvalue of $A$ in
absolute value. The \emph{spectral gap}\index{spectral gap} of $G$ is
given by $1-\lambda_G$.  Starting from a probability distribution $p$
on the set of vertices, represented as a real vector with coordinates
index by the vertex set, performing a single-step random walk on $G$
leads to the distribution defined by $pA$. The following is a well
known lemma on the convergence of the distributions resulting from
random walks (see \cite{ref:Lovasz} for a proof):

%
%
\begin{lem} \label{lem:walk} Let $G=(V,E)$ be a $d$-regular undirected
  graph, and $A$ be its normalized adjacency matrix. Then for any
  probability vector $p$, we have $\| pA - \U_V \|_2 \leq \lambda_G \|
  p - \U_V \|_2$, where $\| \cdot \|_2$ denotes the $\ell_2$
  norm. \qed
\end{lem}

\subsection{The construction and analysis}

The extractor of Kamp and Zuckerman~\cite{ref:KZ}\index{extractor!Kamp
  and Zuckerman's} starts with a fixed vertex in a large expander
graph and interprets the input as the description of a walk on the
graph.  Then it outputs the label of the vertex reached at the end of
the walk.  Notice that a direct approach to invert this function 
amounts to sampling a path of a particular length between a pair of
vertices in the graph, uniformly among all the possibilities, which
might be a difficult problem for good families of expander
graphs\footnote{In fact intractability of the easier problem of
  finding a loop in certain families of expander graphs forms the
  underlying basis for a class of cryptographic hash functions (cf.\
  \cite{ref:CGL}). Even though this easier problem has been solved in
  \cite{ref:TZ08}, uniform sampling of paths seems to be much more
  difficult.}.  We work around this problem by choosing the starting
point of the walk from the input\footnote{ The idea of choosing the
  starting point of the walk from the input sequence has been used
  before in extractor constructions \cite{ref:Zuc07}, but in the
  context of seeded extractors for general sources with high
  entropy.}. The price that we pay by doing so is a slightly 
  larger error compared to the original construction of Kamp and
  Zuckerman that is, asymptotically, of little significance. 
  In particular we show the following:

\begin{thm} \label{thm:invWalk} Let $G$ be a constructible $d$-regular
  graph with $d^m$ vertices and second largest eigenvalue
  $\lambda_G\ge 1/\sqrt{d}$.  Then there exists an explicit invertible
  $(k,2^{s/2})_d$ symbol-fixing extractor $\kz\colon [d]^n \to [d]^m$,
  such that
  \[
  s := \left\{ \begin{array}{ll}
      m \log d + k \log \lambda_G^2 & \text{if $k \leq n - m$,} \\
      (n-k) \log d + (n-m) \log \lambda_G^2 & \text{if $k > n - m$.} \\
    \end{array} \right.
  \]
\end{thm}
\newcommand{\inv}{\mathsf{Inv}}
\begin{Proof}
  We first describe the extractor and its inverse. Given an input $(v,
  w) \in [d]^m \times [d]^{n-m},$ the function $\kz$ interprets $v$ as
  a vertex of $G$ and $w$ as the description of a walk starting from
  $v$. The output is the index of the vertex reached at the end of the
  walk.  Fig.~\ref{fig:graphWalk} depicts the procedure. The
  $4$-regular graph shown in this toy example has $8$ vertices labeled
  with binary sequences of length $3$. Edges of the graph are
  consistently labeled at both endpoints with the set of labels
  $\{1,2,3,4\}$. The input sequence $(0,1,0\mid2,3,4,2,4$) shown below
  the graph describes a walk starting from the vertex $010$ and
  following the path shown by the solid arrows. The output of the
  extractor is the label of the final vertex $011$.

\begin{figure}
\centerline{\includegraphics[width=7cm]{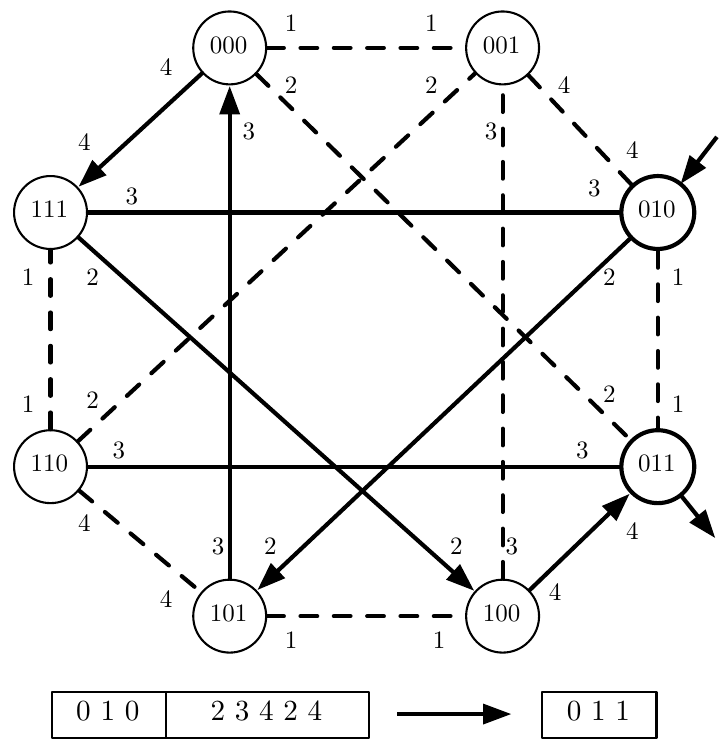}}
\caption[The random-walk symbol-fixing extractor]{The random-walk symbol-fixing extractor.}
\label{fig:graphWalk}
\end{figure}

  The inverter $\inv$ works as follows: Given an input $x \in [d]^m$,
  $x$ is interpreted as a vertex of $G$. Then $\inv$ picks $W \in
  [d]^{n-m}$ uniformly at random.  Let $V$ be the unique vertex starting from
  which the walk described by $W$ ends up in $x$. The inverter outputs
  $(V, W)$. It is easy to verify that $\inv$ satisfies the properties
  of a $0$-inverter.
  
  Now we show that $\kz$ is an extractor with the given parameters. We
  will follow the same line of argument as in the original proof of
  Kamp and Zuckerman.  Let $(v, w) \in [d]^{m} \times [d]^{n-m}$ be a
  vector sampled from an $(n, k)_d$ symbol-fixing source, and let $u
  := \kz(v, w)$.  Recall that $u$ can be seen as the vertex of $G$
  reached at the end of the walk described by $w$ starting from $v$.
  Let $p_i$ denote the probability vector corresponding to the walk
  right after the $i$th step, for $i = 0, \ldots, n-m$, and denote by
  $p$ the uniform probability vector on the vertices of $G$.  Our goal
  is to bound the error $\eps$ of the extractor, which is half the
  $\ell_1$ norm of $p_{n-m} - p$.

  Suppose that $v$ contains $k_1$ random symbols and the remaining
  $k_2 := k - k_1$ random symbols are in $w$. Then $p_0$ has the value
  $d^{-k_1}$ at $d^{k_1}$ of the coordinates and zeros elsewhere,
  hence
  \begin{eqnarray*}
  \| p_0 - p \|_2^2 &=& d^{k_1} (d^{-k_1} - d^{-m})^2 + (d^m - d^{k_1})
  d^{-2m} \\ &=& d^{-k_1} - d^{-m} \leq d^{-k_1}.
  \end{eqnarray*}

  Now for each $i \in [n-m]$, if the $i$th step of the walk
  corresponds to a random symbol in $w$ the $\ell_2$ distance is
  multiplied by $\lambda_G$ by Lemma~\ref{lem:walk}.  Otherwise the
  distance remains the same due to the fact that the labeling of $G$
  is consistent.  Hence we obtain $\| p_{n-m} - p \|_2^2 \leq d^{-k_1}
  \lambda_G^{2 k_2}$.  Translating this into the $\ell_1$ norm by
  using the Cauchy-Schwarz inequality, we obtain $\eps$, namely,
  \begin{equation*} \label{eqn:epsBound} \eps \leq \frac{1}{2}
    d^{(m-k_1)/2} \lambda_G^{k_2} < 2^{((m-k_1) \log d + k_2 \log
      \lambda_G^2)/2}.
  \end{equation*}
  By our assumption, $\lambda_G \geq 1/\sqrt{d}$. Hence, everything
  but $k_1$ and $k_2$ being fixed, the above bound is maximized when
  $k_1$ is minimized.  When $k \leq n-m$, this corresponds to the case
  $k_1 = 0$, and otherwise to the case $k_1 = k-n+m$.  This gives us
  the desired upper bound on $\eps$.
\end{Proof}

Combining this with Theorem~\ref{thm:protocol} and setting up the the
right asymptotic parameters, we obtain our protocol for the wiretap
channel problem.

\newcommand{\pslk}{\rho} 

\begin{coro} \label{coro:WTWalk} Let $\delta \in [0, 1)$ be a constant, 
and suppose that there is a constructible family
  of $d$-regular expander graphs with spectral gap at least
  $1-\lambda$ that is $c$-dense, for constants $\lambda < 1$ and $c
  \geq 1$.

  Then, for every large enough $n$, and arbitrarily small constant $\pslk >
  0$, there is a $(\delta n,
  2^{-\Omega(n)}, 0)_d$-resilient wiretap protocol with block length
  $n$ and rate 
  \begin{equation} \label{eqn:rateWalk}
  R = \left\{ \begin{array}{ll}
  1-\delta/\alpha - \pslk  & \text{if $\delta<\alpha(1-\rho)/(1+\alpha)$,} \\
  \alpha/(1+\alpha) - \pslk & \text{if $\frac{\alpha(1-\rho)}{1+\alpha} \leq \delta < \frac{\alpha}{1+\alpha}$,} \\
  \alpha (1-\delta) - \pslk  & \text{if $\delta \geq \alpha/(1+\alpha)$,} \\   
  \end{array} \right.
  \end{equation}
%
  where $\alpha := -\log_d \lambda^2$.
\end{coro}

\begin{Proof} 
For the case $c = 1$ we use Theorem~\ref{thm:protocol}
  with the extractor $\kz$ of Theorem~\ref{thm:invWalk} and its
  inverse.  Every infinite family of graphs must satisfy $\lambda \geq
  2 \sqrt{d-1}/d$ \cite{ref:nilli}, and in particular we have $\lambda
  \geq 1/\sqrt{d}$, as required by Theorem~\ref{thm:invWalk} (and $\alpha \leq 1$). 
  
  First, we exclude the ``intermediate'' interval   
  $\alpha(1-\rho)/(1+\alpha) \leq \delta < \alpha/(1+\alpha)$ by observing
  that we can always increase $\delta$ (at the cost of a lower rate) so that
  its value falls outside the interval. In particular, in this case,
  we can increase $\delta$ to $\bar{\delta} := \alpha/(1+\alpha)$ and apply the result proved
  later to get a rate $\alpha (1-\bar{\delta}) - \pslk = \alpha/(1+\alpha) - \pslk$.
  Note that the intermediate interval can be made arbitrarily small by
  choosing a sufficiently small value for $\pslk$.
  
  For the remaining values of $\delta$, 
  we choose the parameters $k := (1-\delta) n$ and $m := nR$, where the
  rate $R$ is chosen according to \eqref{eqn:rateWalk}. 
  The parameter $s$ in Theorem~\ref{thm:invWalk} can be rewritten as
  \begin{eqnarray*}
  s &=& \min\{m, n-k\} \log d - \alpha \min\{k, n-m\} \log d \\
  &=& \min\{m, \delta n\} \log d - \alpha (n- \max\{m, \delta n\}) \log d.
  \end{eqnarray*}
  Now consider two cases:
  \begin{enumerate}
  \item $\delta<\alpha(1-\rho)/(1+\alpha)$: In this case, we have
  $R > \delta$ and thus, the value of $s$ simplifies to
 \begin{eqnarray*}
 s &=& (\delta - \alpha + \alpha R) n \log d \\
 &=& (\delta - \alpha + \alpha (1-\delta/\alpha - \pslk)) n \log d \\
 &=&  -\alpha \pslk n \log d  = -\Omega(n).
 \end{eqnarray*}
 \item $\delta \geq \alpha/(1+\alpha)$: In this case, $R \leq \delta$,
 and $s$ can be written as
 \begin{eqnarray*}
 s &=& (R - \alpha + \alpha \delta) n \log d \\
 &=& (\alpha(1-\delta) - \rho - \alpha + \alpha \delta) n \log d \\
 &=&  -\pslk n \log d  = -\Omega(n).
 \end{eqnarray*}
  \end{enumerate}
  Hence, we always have $s = -\Omega(n)$ and exponentially small error.  The case $c > 1$
  is similar, but involves technicalities for dealing with lack of
  graphs of arbitrary size in the family. We will elaborate on this in
  Appendix~\ref{app:wiretapDetails}.
\end{Proof}

Using explicit constructions of Ramanujan graphs that achieve \[\lambda
\leq 2 \sqrt{d-1}/d\] when $d-1$ is a prime power
\cites{ref:ramanujan1,ref:ramanujan2,ref:pizer}, one can obtain
$\alpha \geq 1 - 2/\log d$, which can be made arbitrarily close to one
(hence, making the protocol arbitrarily close to the optimal bound) by
choosing a suitable alphabet size that does {not} depend on $n$. Namely, we have
the following result:

\begin{coro} \label{coro:WTWalkRamanujan} Let $\delta \in [0, 1)$ and $\pslk >
  0$ be arbitrary constants. Then, there is a positive integer $d$ only depending
  on $\pslk$ such that the following holds:
  For every large enough $n$, there is a $(\delta n,
  2^{-\Omega(n)}, 0)_d$-resilient wiretap protocol with block length
  $n$ and rate at least $1-\delta - \pslk$. \qed
\end{coro}

\section[Invertible Affine Extractors]{Invertible Affine Extractors
  and Asymptotically Optimal Wiretap Protocols}
\label{sec:invAExt}
In this section we will construct a black box transformation for
making certain seedless extractors invertible.  The method is
described in detail for affine extractors, and leads to wiretap
protocols with asymptotically optimal rate-resilience trade-offs.
Being based on affine extractors, these protocols
are only defined for prime power alphabet sizes. On the other hand,
the random-walk based protocol discussed in Section~\ref{sec:walk}
can be potentially instantiated for an arbitrary alphabet size, though
achieving asymptotically sub-optimal parameters (and a positive
rate only for an alphabet of size $3$ or more).

Modulo some minor differences, the construction can be simply
described as follows: A seedless affine extractor is first used to
extract a small number of uniform random bits from the source, and the
resulting sequence is then used as the seed for a seeded extractor
that extracts almost the entire entropy of the source.

Of course, seeded extractors in general are not guaranteed to work if
(as in the above construction) their seed is not independent from the
source. However, as observed by Gabizon and Raz~\cite{ref:affine}, a
\emph{linear} seeded extractor can extract from an affine source if
the seed is the outcome of an affine extractor on the source. This
idea was formalized in a more general setting by
Shaltiel~\cite{ref:mileage}.

Shaltiel's result gives a general framework for transforming any
seedless extractor (for a family of sources satisfying a certain
\emph{closedness} condition) with short output length to one with an
almost optimal output length. The construction uses the imperfect
seedless extractor to extract a small number of uniform random bits
from the source, and will then use the resulting sequence as the seed
for a seeded extractor to extract more random bits from the
source. For a suitable choice of the seeded extractor, one can use
this construction to extract almost all min-entropy of the source.

The closedness condition needed for this result to work for a family
$\cC$ of sources is that, letting $\extf(x, z)$ denote the seeded
extractor with seed $z$, for every $\cX \in \cC$ and every fixed $z$
and $y$, the distribution $(\cX | \extf(\cX, z)=y)$ belongs to $\cC$. If
$\extf$ is a linear function for every fixed $s$, the result will be
available for affine sources (since we are imposing a linear
constraint on an affine source, it remains an affine source).  A more
precise statement of Shaltiel's main result is the following:

\begin{thm} \cite{ref:mileage} \label{thm:mileage} Let $\cC$ be a
  class of distributions on $\zo^n$ and $\extf\colon \zo^n \to \zo^t$
  be an extractor for $\cC$ with error $\eps$. Let $F\colon \zo^n
  \times \zo^t \to \zo^m$ be a function for which $\cC$ satisfies
  the closedness condition above. Then for every $\cX \in \cC$,
  $F(\cX, \extf(\cX)) \sim_{\eps \cdot 2^{t+3}} F(\cX, \U_t)$. \qed
\end{thm}

A seeded extractor is called \emph{linear} if it is a
linear function for every fixed choice of the seed. This
condition is in particular satisfied by Trevisan's extractor~\cite{ref:Tre}.  For
our construction, we will use the following 
improvement of this extractor due to Raz, Reingold and Vadhan \cite{ref:RRV}:

\begin{thm} \cite{ref:RRV} \label{thm:seeded} There is an explicit
  strong linear seeded $(k, \eps)$-extractor $\extr\colon \F_2^n
  \times \zo^r \to \F_2^m$ with $r = O(\log^3 (n/\eps))$ and $m = k -
  O(r)$. \qed
\end{thm}

\begin{rem}
  We note that our arguments would identically work for any other
  linear seeded extractor as well, for instance those constructed in
  \cites{ref:RMExtractor,ref:SU}.  However, the most crucial parameter
  in our application is the output length of the extractor, being
  closely related to the rate of the wiretap protocols we
  obtain. Among the constructions we are aware of, the result quoted
  in Theorem~\ref{thm:seeded} is the best in this regard. Moreover, an
  affine seeded extractor with better parameters is constructed by
  Gabizon and Raz~\cite{ref:affine}, but it requires a large alphabet
  size to work.
\end{rem}

Now, having the right tools in hand, we are ready to formally describe
our construction of invertible affine extractors with nearly optimal
output length. Broadly speaking, the construction follows the abovementioned idea
of Shaltiel, Gabizon, and Raz \cites{ref:mileage,ref:affine} on enlarging
the output length of affine extractors, with an additional
``twist'' for making the extractor invertible. For concreteness, the description is given over the
binary field $\F_2$:

\begin{thm} \label{thm:invAffine} For every constant $\delta \in (0,
  1]$ and every $\alpha \in (0,1)$, there is an explicit $0$-invertible
  affine extractor $\extf\colon \F_2^n \to \F_2^m$ for min-entropy $\delta
  n$ with output length $m = \delta n - O(n^\alpha)$ and error 
  $\eps = O(2^{-n^{\alpha/3}})$.
\end{thm}

\begin{Proof} 
  Let $\eps := 2^{-n^{\alpha/3}}$, and $t := O(\log^3 (n/\eps)) =
  O(n^\alpha)$ be the seed length required by the extractor $\extr$ in
  Theorem~\ref{thm:seeded} for input length $n$ and error $\eps$, and
  further, let $n' := n - t$. Set up $\extr$ for input length $n'$,
  min-entropy $\delta n - t$, seed length $t$ and error $\eps$.  Also
  set up Bourgain's extractor $\bou$ (Theorem~\ref{thm:Bourgain}) for input length $n'$ and entropy
  rate $\delta'$, for an arbitrary constant $\delta' < \delta$.  Then
  the function $\extf$ will view the $n$-bit input sequence as a tuple
  $(s, x)$, $s \in \F_2^t$ and $x \in \F_2^{n'}$, and outputs
  $\extr(x, s + \bou(x)|_{[t]})$.  This is depicted in Fig.~\ref{fig:affineExt}.

\begin{figure*}
\centerline{\includegraphics[width=10cm]{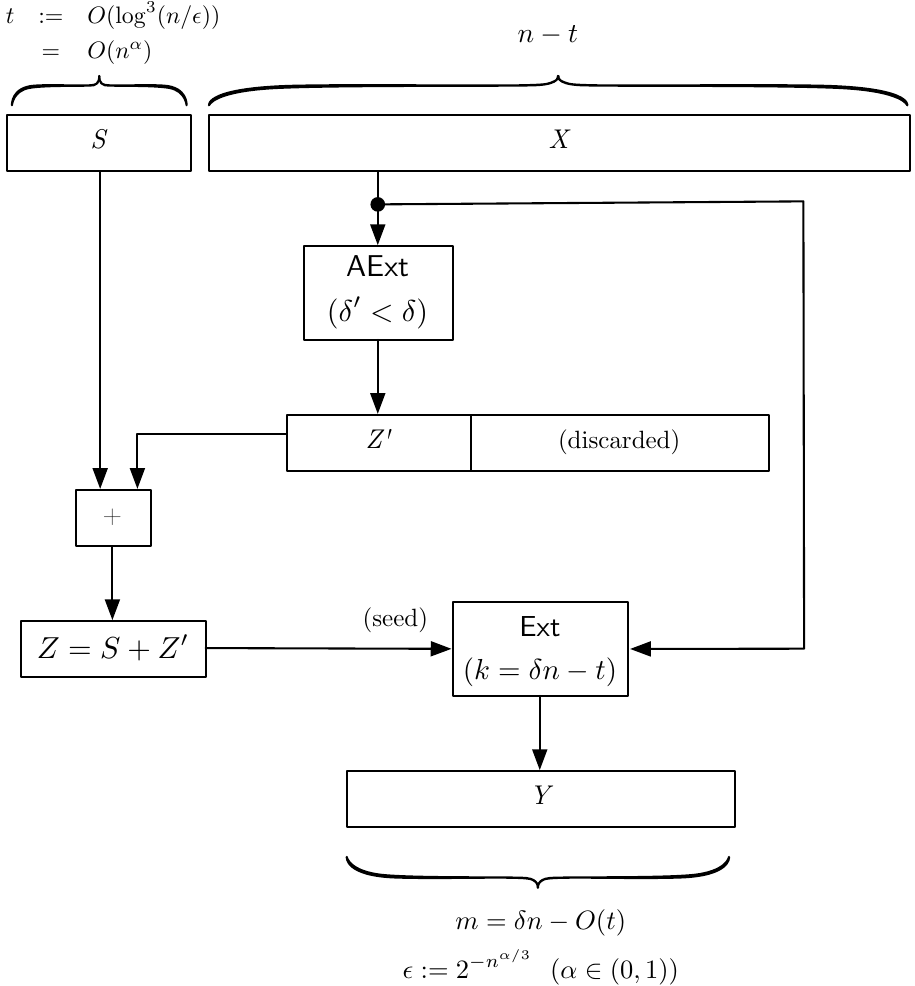}}
\caption[Construction of the invertible affine extractor]{Construction of the invertible affine extractor.}
\label{fig:affineExt}
\end{figure*}

  First we show that the construction gives an affine extractor. Suppose that $(S, X)
  \in \F_2^t \times \F_2^{n'}$ is a random variable sampled from an
  affine distribution with min-entropy $\delta n$. The variable $S$
  can have an affine dependency on $X$.  Hence, for every fixed $s \in
  \F_2^t$, the distribution of $X$ conditioned on the event $S=s$ is
  affine with min-entropy at least $\delta n - t$, which is at least
  $\delta' n'$ for large enough $n$. Hence $\bou(X)$ will be
  $2^{-\Omega(n)}$-close to uniform by Theorem~\ref{thm:Bourgain}.
  This implies that $\bou(X)|_{[t]} + S$ can extract $t$ random bits
  from the affine source with error $2^{-\Omega(n)}$.  Combining this
  with Theorem~\ref{thm:mileage}, and noticing the fact that the class of
  affine extractors is closed with respect to linear seeded
  extractors, we conclude that $\extf$ is an affine extractor with error
  at most $\eps + 2^{-\Omega(n)} \cdot 2^{t+3} = O(2^{-n^{\alpha /
      3}})$.

  Now the inverter works as follows: Given $y \in \F_2^m$, first it
  picks $Z \in \F_2^t$ uniformly at random. The seeded extractor
  $\extr$, given the seed $Z$ is a linear function $\extr_Z\colon
  \F_2^{n'} \to \F_2^m$.  Without loss of generality, assume that this
  function is surjective\footnote{Because the seeded extractor is
    strong and linear, for most choices of the seed it is a good
    extractor (by an averaging argument), 
    and hence necessarily surjective (if not, one of the
    output symbols would linearly depend on the others and obviously
    the output distribution would not be close to uniform).  Hence if $\extr$ is
    not surjective for some seed $z$, one can replace it by a trivial
    surjective linear mapping without affecting its extraction
    properties.}.  Then the inverter picks $X \in \F_2^{n'}$ uniformly
  at random from the affine subspace defined by the linear constraint
  $\extr_Z(X) = y$, and outputs $(Z + \bou(X)|_{[t]}, X)$. It is easy
  to verify that the output is indeed a valid preimage of $y$. To see
  the uniformity of the inverter, note that if $y$ is chosen uniformly
  at random, the distribution of $(Z, X)$ will be uniform on
  $\F_2^n$. Hence $(Z + \bou(X)|_{[t]}, X)$, which is the output of
  the inverter, will be uniform.
\end{Proof}

In the above construction we are using an affine and a linear seeded
extractor as \emph{black boxes}, and hence, they can be replaced by
any other extractors as well (the construction will achieve an optimal
rate provided that the seeded extractor extracts almost the entire
source entropy).  In particular, over large fields one can use the
affine and seeded extractors given by Gabizon and Raz
\cite{ref:affine} that work for sub-constant entropy rates as well.
  
Moreover, for concreteness we described and instantiated our
construction over the binary field. Observe that Shaltiel's result,
for the special case of affine sources, holds regardless of the
alphabet size. Moreover, Trevisan's linear seeded extractor can be
naturally extended to handle arbitrary alphabets. Hence, in order to
extend our result to non-binary alphabets, it suffices to ensure that
a suitable seedless affine extractor that supports the desired
alphabet size is available. Bourgain's original
result~\cite{ref:Bourgain} is stated and proved for the binary
alphabet; however, this result can be adapted to work
for larger fields as well~\cite{ref:BourPriv}. Such an extension
(along with some improvements and simplifications) is made explicit by
Yehudayoff \cite{ref:Yeh09}.

An affine extractor is in particular, a symbol-fixing extractor.
Hence Theorem~\ref{thm:invAffine}, combined with
Theorem~\ref{thm:protocol} gives us a wiretap protocol with almost
optimal parameters:

\begin{thm} \label{coro:wiretap} Let $\delta \in [0, 1)$ and $\alpha
  \in (0, 1/3)$ be constants.  Then for a prime power $q > 1$ and
  every large enough $n$ there is a $(\delta n, O(2^{-n^{\alpha}}),
  0)_q$-resilient wiretap protocol with block length $n$ and rate $1 -
  \delta - o(1)$. \qed
\end{thm}

\section{Further Applications and Extensions}
\label{sec:Apps}
In this section we will sketch some important applications of our
technique to more general wiretap problems.

\subsection{Noisy Channels and Active Intruders}
\label{sec:activeIntruder}
Suppose that Alice wants to transmit a particular sequence to Bob
through a noisy channel. She can use various techniques from coding
theory to encode her information and protect it against noise. Now
what if there is an intruder who can partially observe the transmitted
sequence and even {manipulate} it?  Modification of the sequence by
the intruder can be regarded in the same way as the channel noise;
thus one gets security against active intrusion as a ``{bonus}'' by
constructing a code that is resilient against noise and {passive}
eavesdropping.  There are two natural and {modular} approaches to
construct such a code.

A possible attempt would be to first encode the message using a good
error-correcting code and then to apply a wiretap encoder to protect
the encoded sequence against the wiretapper. However, this will not
necessarily keep the information protected against the channel noise,
as the combination of the wiretap encoder and decoder does not have to
be resistant to noise.

Another attempt is to first use a wiretap encoder and then apply an
error-correcting code on the resulting sequence. Here it is not
necessarily the case that the information will be kept secure against
intrusion anymore, as the wiretapper now gets to observe the bits from
the channel-encoded sequence that may reveal information about the
original sequence. However, the wiretap protocol given in
Theorem~\ref{coro:wiretap} is constructed from an invertible {affine}
extractor, and guarantees resiliency even if the intruder is allowed
to observe arbitrary {linear combinations} of the transmitted
sequence (in this case, the
distribution of the encoded sequence subject to the intruder's observation
becomes an affine source and thus, the arguments of the proof of Theorem~\ref{thm:protocol}
remain valid). In particular, Theorem~\ref{coro:wiretap} holds even
if the intruder's observation is allowed to be obtained after applying any
arbitrary linear mapping on the output of the wiretap encoder. 
Hence, we can use the wiretap scheme as an outer code and 
still ensure privacy against an active intruder and reliability in
presence of a noisy channel, provided that the
error-correcting code being used as the inner code is linear.  This immediately gives us the
following result:

\begin{thm} \label{thm:combined}
  Suppose that there is a $q$-ary linear error-correcting code with
  rate $R$ that is able to correct up to a $\tau$ fraction of errors
  (via unique or list decoding). Then for every constant $\delta \in
  [0, 1)$ and $\alpha \in (0,1/3)$ and large enough $n$, there is a
  $(\delta n, O(2^{-n^{\alpha}}), 0)_q$-resilient wiretap protocol
  with block length $n$ and rate $R - \delta - o(1)$ that can also
  correct up to a $\tau$ fraction of errors. 
\end{thm}

\begin{proof}
 We use the wiretap code given by Theorem~\ref{coro:wiretap} as the outer code
 and the linear channel code with rate $R$ as the inner code. 
 Suppose that the rate of the outer code is $R'$ and the block length of
 the combined code is $n$. 
 By the discussion above and using Theorem~\ref{coro:wiretap}, the combined code tolerates against 
 $Rn(1-R'-o(1))$ bits of observation by the intruder, even after linear post-processings
 of the inner code. Therefore the resilience of the combined code can be written as
 \[
  \delta = R(1-R'-o(1)),
 \]
 from which we can compute the rate of the combined code as
 \[
  R R' = R - \delta - o(1).
 \]
\end{proof}

\begin{figure*}
\centerline{\includegraphics[width=0.8\textwidth]{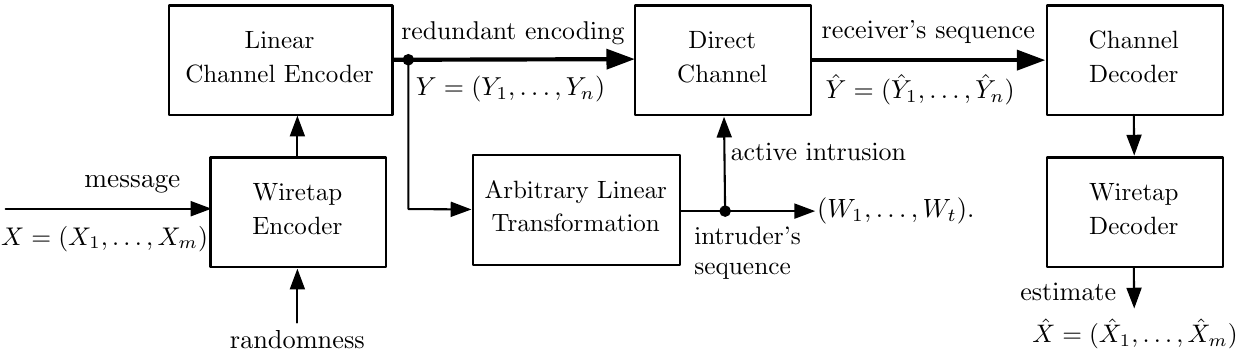}}
\caption[Wiretap scheme composed with channel coding]
{Wiretap scheme composed with channel coding. If the wiretap scheme is
constructed by an invertible affine extractor, it can guarantee secrecy
even in presence of arbitrary linear manipulation of the information.
Active intrusion can be defied using an error-correcting inner code.}
\label{fig:wiretapNoise}
\end{figure*}

The setting discussed above is shown in Fig.~\ref{fig:wiretapNoise}.
The same idea can be used to protect fountain codes, e.g.,
LT-~\cite{ref:LT} and Raptor Codes \cite{ref:Raptor}, against
wiretappers without affecting the error correction capabilities of the
code.
  
Obviously this simple composition idea can be used for any type of
channel so long as the inner code is linear, at the cost of reducing
the total rate by almost $\delta$ (following exactly the same argument
as in Theorem~\ref{thm:combined}). Hence, if the inner code achieves
the Shannon capacity of the direct channel (in the absence of the
wiretapper), the composed code will have a rate that is smaller than the
direct channel capacity by the resilience parameter $\delta$. 
This is known to be the best possible rate when the channels
are discrete, memoryless, and symmetric and the wiretap channel is 
a \emph{degraded} version of the symmetric channel \cites{ref:CK78,ref:LYC77}.

\subsection{Network Coding}
\label{sec:NetCod}
Our wiretap protocol from invertible affine extractors is also
applicable in the more general setting of transmission over
{networks}.  In this work we focus on a network setting known as 
\emph{multicast}. A multicast communication network can be modeled as a directed
graph, in which nodes represent the network devices and information is
transmitted along the edges. One particular node is identified as the
\emph{source} and $m$ nodes are identified as \emph{receivers}. The
main problem in network coding is to have the source reliably transmit
information to the receivers at the highest possible rate, while
allowing the intermediate nodes arbitrarily process the information
along the way.

Suppose that, in the graph that defines the topology of the network,
the min-cut between the source to each receiver is $n$. It was shown
in \cite{ref:NetCod1} that the source can transmit information up to
rate $n$ (symbols per transmission) to all receivers (which is
optimal), and in \cites{ref:NetCod2,ref:NetCod3} that {linear} network
coding is in fact sufficient to achieve this rate. That is, the
transmission at rate $n$ is possible when the intermediate nodes are
allowed to forward packets that are (as symbols over a finite field)
linear combinations of the packets that they receive (See
\cite{ref:NetCodBook} for a comprehensive account of these and other
relevant results).

\begin{figure*}
\centerline{\includegraphics[width=0.8\textwidth]{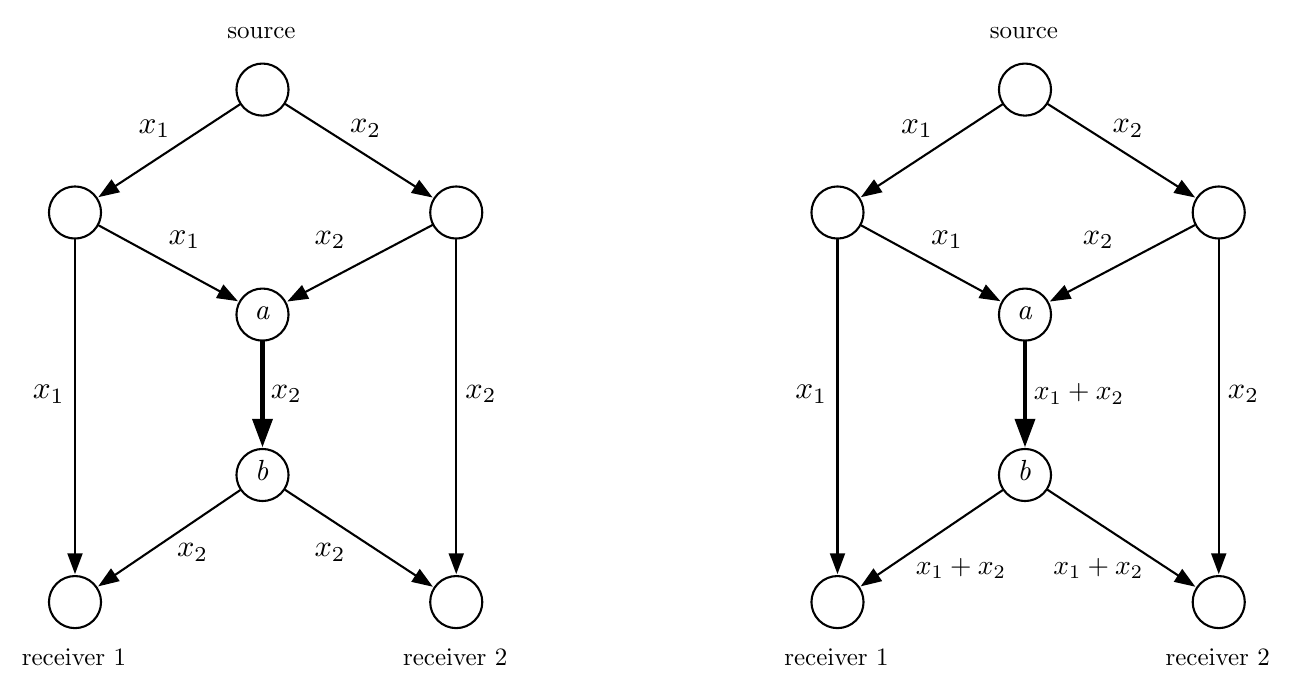}}
\caption[Network coding versus unprocessed forwarding]
{Network coding (right), versus unprocessed forwarding (left).}
\label{fig:netcod}
\end{figure*}

A basic example is shown by the \emph{butterfly} network in
Fig.~\ref{fig:netcod}.  This network consists of a source on the top
and two receivers on the bottom, where the min-cut to each receiver is~$2$. 
Without processing the incoming data, as in the left figure, one
of the two receivers may receive information at the optimal rate of~$2$ 
symbols per transmission (namely, receiver~$1$ in the figure).
However, due to the bottleneck existing in the middle (shown by the
thick edge $a\to b$), the other receiver will be forced to receive at
an inferior rate of~$1$ symbol per transmission. However, if linear
processing of the information is allowed, node $a$ may combine its
incoming information by treating packets as symbols over a finite
field and adding them up, as in the right figure. Both receivers may
then solve a full-rank system of linear equations to retrieve the
original source symbols $x_1$ and $x_2$, and thereby achieve the optimal
min-cut rate.

Designing wiretap protocols for networks is an important question in
network coding, which was first posed by Cai and Yeung
\cite{ref:NetWT}. In this problem, an intruder can choose a bounded
number, say $t$, of the edges and eavesdrop all the packets going
through those edges. They designed a network code that could provide
the optimal multicast rate of $n - t$ with perfect privacy. However
this code requires an alphabet size of order $\binom{|E|}{t}$, where
$E$ is the set of edges.  Their result was later improved in
\cite{ref:FMSS04} who showed that a random linear coding scheme can
provide privacy with a much smaller alphabet size if one is willing to
achieve a slightly sub-optimal rate. Namely, they obtain rate
$n-t(1+\eps)$ with an alphabet of size roughly
$\mathrm{\Theta}(|E|^{1/\eps})$, and show that achieving the exact
optimal rate is not possible with small alphabet size.

El~Rouayheb and Soljanin \cite{ref:Emina} suggested to use the
original code of Ozarow and Wyner \cite{ref:Wyner2} as an {outer code}
at the source and showed that a careful choice of the network code can
provide optimal rate with perfect privacy. However, their code
eventually needs an alphabet of size at least $\binom{|E|-1}{t-1} +
m$. Building upon this work, Silva and Kschischang
\cite{ref:RankMetric} constructed an outer code that provides similar
results while leaving the underlying network code unchanged.  However,
their result comes at the cost of increasing the packet size by a
multiplicative factor of at least the min-cut bound, $n$ (or in
mathematical terms, the original alphabet size $q$ of the network is
enlarged to at least $q^n$).  For practical purposes, this is an
acceptable solution provided that an estimate on the min-cut size of
the network is available at the wiretap encoder.

\begin{figure}
\centerline{\includegraphics[width=8cm]{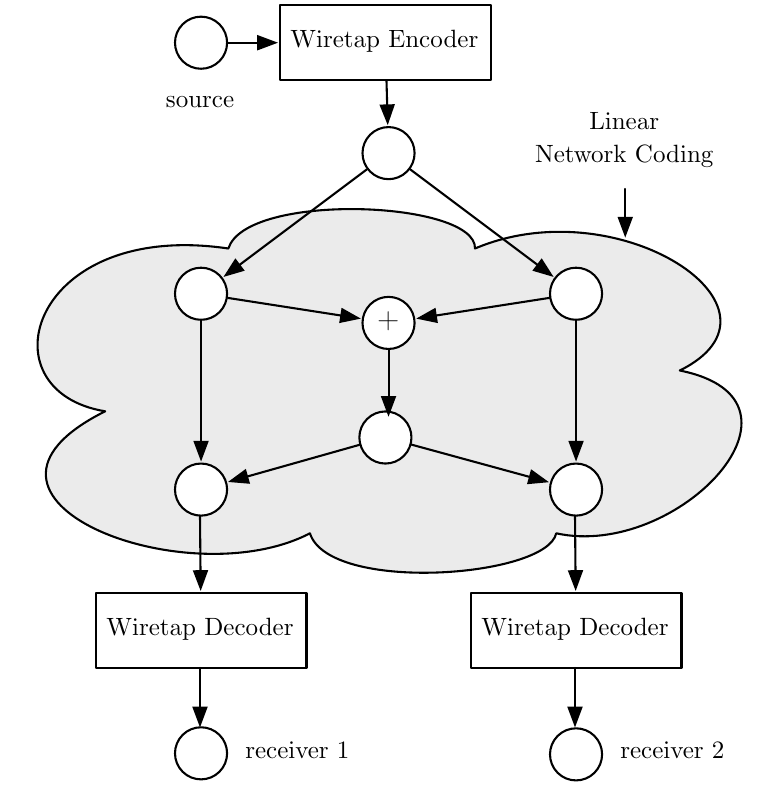}}
\caption[Linear network coding with an outer layer of wiretap
  encoding added for providing secrecy]{Linear network coding with an outer layer of wiretap
  encoding added for providing secrecy.}
\label{fig:netCodWiretap}
\end{figure}

By the discussion presented in Section~\ref{sec:activeIntruder}, 
the rate-optimal wiretap protocol given in Theorem~\ref{coro:wiretap}
stays resilient even in presence of any linear post-processing of the
encoded information. Thus, using the wiretap encoder given by this
result as an outer-code in the source node, one can construct an
asymptotically optimal wiretap protocol for networks that is
completely unaware of the network and eliminates all the restrictions
in the above results. This is schematically shown in Fig.~\ref{fig:netCodWiretap}. 
Hence, extending our notion of $(t, \eps,
\gamma)_q$-resilient wiretap protocols naturally to communication
networks, we obtain the following:

\begin{thm}
  Let $\delta \in [0, 1)$ and $\alpha \in (0, 1/3)$ be constants, and
  consider a network that uses a linear coding scheme over a finite
  field $\F_{q}$ for reliably transmitting information at rate $R$.
Suppose that, at each transmission, an intruder can arbitrarily observe
up to $\delta R$ intermediate links in the network. 
Then the source
  and the receiver nodes can use an outer code of rate $1 - \delta -
  o(1)$ (obtaining a total rate of $R(1-\delta) - o(1)$) which is completely 
  independent of the network, leaves the
  network code unchanged, and provides almost perfect privacy with error
  $O(2^{-{R}^{\alpha}})$ and zero leakage over a $q$-ary
  alphabet. \qed
\end{thm}

In addition to the above result that uses the invertible affine extractor
of Theorem~\ref{thm:invAffine}, it is possible to use other rate-optimal
invertiable affine extractors. In particular,
observe that the restricted affine extractor of
Theorem~\ref{thm:resAffExtr} (and in particular, Corollary~\ref{coro:resAffExtrGabidulin}) 
is a linear function (over the extension
field) and is thus, obviously has an efficient $0$-inverter (since
inverting the extractor amounts to solving a system of linear
equations).  By using this extractor (instantiated with Gabidulin's
MRD codes as in Corollary~\ref{coro:resAffExtrGabidulin}),
 we may recover the result of Silva and Kschischang
\cite{ref:RankMetric} in our framework. More precisely, we have the following result:

\begin{coro}
  Let $q$ be any prime power, and
  consider a network with minimum cut of size $n$ that uses a linear coding scheme over 
  $\F_{q}$ for reliably transmitting information at rate $R$.
Suppose that, at each transmission, an intruder can arbitrarily observe up to 
$\delta R$ intermediate links in the network, for some $\delta \in [0, 1)$. 
Then the source
  and the receiver nodes can use an outer code of rate $1 - \delta$ 
  over $\F_{q^n}$ (obtaining a total rate of $R(1-\delta)$) 
  that provides perfect privacy over a $q^n$-ary
  alphabet. \qed
\end{coro}

\subsection{Arbitrary Processing}
\label{subsec:arbitrary}

In this section we consider the erasure wiretap problem in its most
general setting, which is still of practical importance.  Suppose that
the information emitted by the source goes through an arbitrary
communication medium and is arbitrarily processed on the way to
provide protection against noise, to obtain better throughput, or for
other reasons. Now consider an intruder who is able to eavesdrop a
bounded amount of information at various points of the channel. One
can model this scenario in the same way as the original point-to-point
wiretap channel problem (depicted in Fig.~\ref{fig:wiretapII}), 
with the difference that instead of observing
$t$ arbitrarily chosen bits, the intruder now gets to choose an
arbitrary Boolean circuit $\cC$ with $t$ output bits (which captures the
accumulation of all the intermediate processing) and observes the
output of the circuit when applied to the transmitted
sequence\footnote{ In fact this models a ``harder'' problem, as in our
  problem the circuit $\cC$ is given by the communication scheme and
  not the intruder. Nevertheless, we consider the harder problem.}.
  
Obviously there is no way to guarantee resiliency in this setting,
since the intruder can simply choose $\cC$ to compute $t$ output bits
of the wiretap decoder.  However, suppose that in addition there is an
auxiliary communication channel between the source and the receiver
(that we call the \emph{side channel}) that is separated from the main
channel, and hence, the information passed through the two channel do
not \emph{blend} together by the intermediate processing.

\begin{figure}
\centerline{\includegraphics[width=8cm]{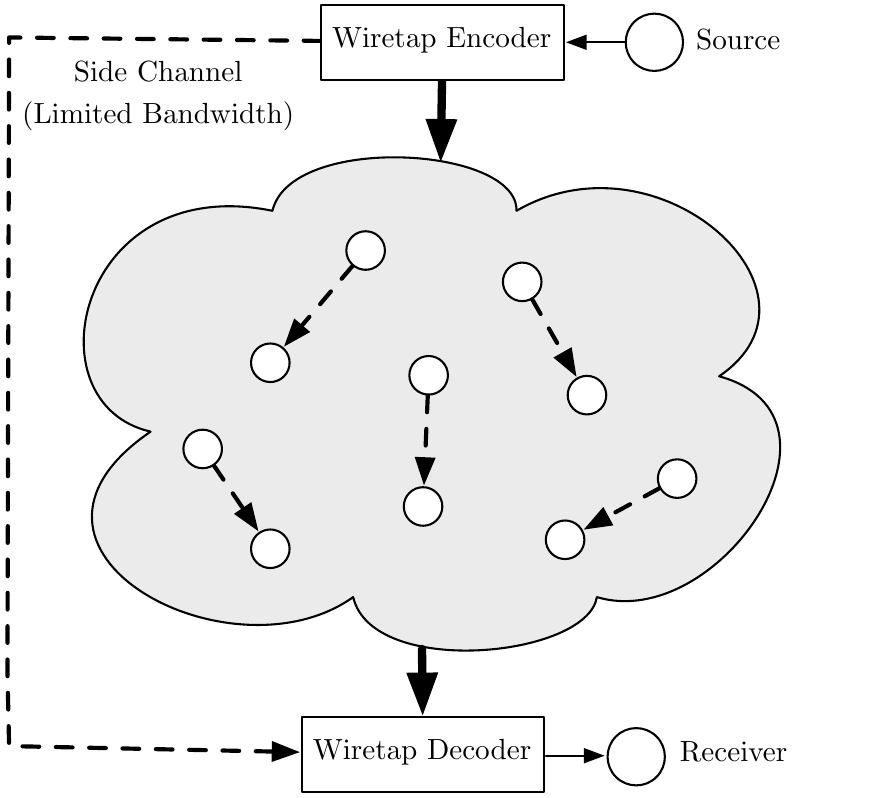}}
\caption[The wiretap channel problem in presence of arbitrary
intermediate processing]{The wiretap channel problem in presence of
  arbitrary intermediate processing. In this example, data is
  transmitted over a packet network (shown as a cloud) in which some
  intermediate links (showed by the dashed arrows), as well as 
  the side channel (also dashed), are accessible to
  an intruder.}
\label{fig:arbitraryWiretap}
\end{figure}

We call this scenario, shown in Fig.~\ref{fig:arbitraryWiretap}, 
the \emph{general wiretap problem} and extend
our notion of $(t, \eps, \gamma)$-resilient protocol to this problem.
The slight modification in the generalized notion is 
that now the output of the encoder (and
the input of the decoder) is a pair of strings $(Y_1, Y_2) \in \zo^n
\times \zo^{n'}$, where $Y_1$ (resp., $Y_2$) is sent through the main
(resp., side) channel.  Now we call $n+n'$ the block length and let the
intruder choose an arbitrary Boolean function $\cC\colon \zo^n \to \zo^t$, 
and observe $(\cC(Y_1), Y_2)$.

More precisely, the generalized model that we consider can be described
as follows:

\begin{enumerate}
\item A wiretap encoder $E(X,Z)$ encodes a message $X \in \zo^m$ using a 
uniformly random seed $Z$. The output of the encoder is a pair of
strings $(Y_1, Y_2)$.

\item The encoded strings $Y_1$ and $Y_2$ are 
sent through the main channel and the side channel, respectively.

\item The intruder chooses an arbitrary function $\cC\colon \zo^n \to \zo^t$, 
and observes $(\cC(Y_1), Y_2)$.

\item A wiretap decoder $D(Y_1, Y_2)$ receives the encoded
strings $(Y_1, Y_2)$ and reconstructs the message $X$.
\end{enumerate}


For a pair of strings $(w_1, w_2) \in \zo^t \times \zo^{n'}$, denote by $\cX_{(w_1, w_2)}$ 
the distribution of the message $X$ conditioned on the event 
$(\cC(Y_1), Y_2) = (w_1, w_2)$.
Now we call a generalized wiretap encoder/decoder pair $(t, \eps, \gamma)$-resilient
if, for every choice of the function $\cC$ chosen by the intruder,
the following holds. Similar to Definition~\ref{def:wiretap},
denote by $B_\cC$ the set of \emph{bad observations}
\[
B_\cC := \{ (w_1, w_2) \in \zo^t \times \zo^{n'}\colon \dist(X, \cX_{(w_1, w_2)}) > \eps \}.
\]
Then the wiretap protocol must satisfy
\[
\Pr[(\cC(Y_1), Y_2) \in B_\cC] \leq \gamma.
\]
The parameters $\eps$ and $\gamma$ are again called the error and the leakage of
the protocol. The rate and resilience parameters are now given by
$R := m/(n+n')$ and $\delta := t/n$, respectively.

The standard definition of wiretap protocols (Definition~\ref{def:wiretap})
corrsponds to the case when $\cC(Y_1) = Y_1|_S$, for some $S \subseteq [n]$,
and $Y_2$ is the empty string. Since the information in $Y_2$, transmitted
through the side channel, does not affect the resilience parameter and
only lowers the rate, we see that 
the infor\-mation-theoretic upper bound for the achievable rates in the
original wiretap problem (namely, $R \leq 1-\delta+o(1)$) extends to the generalized wiretap
problem as well.  Below we show that for the generalized problem, secure
transmission is indeed possible at asymptotically optimal rates (that is,
$R \geq 1-\delta -o(1)$).

%

As before, our idea is to use invertible extractors to
construct general wiretap protocols, but this time we use invertible
strong seeded extractors. Strong seeded extractors were used in
\cite{ref:CDH} to construct ERFs, and this is exactly what we use as
the decoder in our protocol.  As the encoder we will use the
corresponding inverter, which outputs a pair of strings, one for the
extractor's input which is sent through the main channel and another
as the seed which is sent through the side channel.  Hence we will
obtain the following result:

\begin{thm} \label{thm:generalWiretap} Let $\delta \in [0, 1)$ be a
  constant. Then for every $\alpha, \eps > 0$, there is a $(\delta n,
  \eps, 2^{-\alpha n}+\eps)$-resilient wiretap protocol for the
  general wiretap channel problem that sends $n$ bits through
  the main channel and $n' = O(\log^3(n/\eps^2))$ bits through the
  side channel and achieves rate $1 - \delta - \alpha -
  O(n'/(n+n'))$. The protocol is secure even when the entire
  communication through the side channel is observable by the
  intruder.
\end{thm}

\begin{proof} 
  We will need the following claim in our proof, which is easy to
  verify using an averaging argument:
  \begin{claims} \label{prop:preimage} Let $f\colon \zo^n \to
    \zo^{\delta n}$ be any Boolean function. Then for every $\alpha >
    0$, and $X \sim \U_n$, the probability that $f(X)$ has fewer than
    $2^{n(1-\delta-\alpha)}$ preimages is at most $2^{-\alpha n}$.
  \end{claims}
  \begin{proof}[Proof of Claim]
  Define \[ B := \{ y\in \zo^m\colon |f^{-1}(y)| < 2^{n(1- \delta -\alpha)} \}.\]
  Thus, the probability that we wish to bound is 
  \begin{eqnarray*}
  \Pr_X[ f(X) \in B ] &=& 2^{-n} \sum_{y \in B} |f^{-1}(y)| \\ &<&
  2^{-n} \sum_{y \in B} |f^{-1}(y)| 2^{n(1- \delta -\alpha)} < 2^{-\alpha},
  \end{eqnarray*}
  where the last inequality is from the trivial bound $|B| \leq 2^{\delta n}$.
  \end{proof}

  Now, let $\extr$ be the linear seeded extractor of
  Theorem~\ref{thm:seeded}, set up for input length $n$, seed length
  $n' = O(\log^3(n/\eps^2))$, min-entropy $n(1-\delta-\alpha)$, 
  output length $m = n(1-\delta-\alpha) - O(n')$, and error $\eps^2$.
  Then the encoder chooses a seed $Z$ for the extractor uniformly at
  random and sends it through the side channel.
  
  For the chosen value of $Z$, the extractor is a linear function, and
  as before, given a message $X \in \zo^m$, the encoder picks a random
  vector $Y$ in the affine subspace that is mapped by this linear function
  to $X$ and sends it through the public channel. By Remark~\ref{rem:linearInvertible},
  the encoder is polynomial-time computable and $Y$ is uniformly distributed on
  $\zo^n$.
  
  The decoder, upon receiving $(Y,Z)$, applies the extractor to the seed $Z$ received
  through the side channel and the transmitted string $Y$. 
This obviously reproduces the sent message $X = \extr(Y,Z)$.  
  The resiliency
  of the protocol can be shown in a similar manner as in
  Theorem~\ref{thm:protocol}, as follows. Suppose that the intruder
  observes the seed $Z$ and $W := \cC(Y) \in \zo^{\delta n}$, for an
  arbitrary function $\cC\colon \zo^n \to \zo^{\delta n}$.
  
  First, note that by
  the above claim, with probability at least $1-2^{-\alpha n}$, the
  string $Y$ transmitted through the main channel, conditioned on the
  observation $W$ of the intruder from the main channel, has a
  distribution $\cY$ with min-entropy at least
  $n(1-\delta-\alpha)$. Call a particular realization $w \in \zo^{\delta n}$
  of the observation outcome $W$ \emph{good}
  if it makes $\cY$ satisfies this property. In the sequel, we condition
  the random variable $W$ to a good outcome $w$.
  
  Now suppose that the seed $Z$ is 
  entirely revealed to the intruder, and let $z \in \zo^{n'}$ denote the
  particular realization of $Z$. By an averaging argument, with
  probability at least $1-\eps$, $Z$ is a \emph{good} seed for $\cY$,
  in the sense that $\extr(\cY, z)$ is $\eps$-close to uniform. 
  
  Therefore, conditioned on the event that both parts of the observation outcome $(w, z)$
  are good, the message distribution conditioned
  on the intruder's observation (which is given by $\extr(\cY, z)$) is
  $\eps$-close to uniform. The leakage parameter is given by the probability
  that either the part of the observation outcome corresponding to the main channel
  or the part corresponding to the side channel is not good. By a union bound
  this probability is upper bounded by $2^{-\alpha n} + \eps$, which completes the proof
  of resiliency.
\end{proof}



We observe that it is not possible to guarantee zero leakage for the
general wiretap problem above.  As an extreme case, suppose that the function $\cC$ 
is chosen in a way that it has a single preimage for a
particular output $w$ (i.e., $|\cC^{-1}(w)| = 1$).  With nonzero probability the
observation of the intruder from the main channel may turn out to be $w$, in which
case the entire message is revealed (since the intruder learns the entire communication
$(Y_1, Y_2)$ in this case).  Nevertheless, it is possible to
guarantee negligible leakage as the above theorem does.  

Finally, we remark that the general protocol above can be used for the original wiretap~II
problem (where there is no intermediate processing involved). In this case,
both encoded strings $Y_1$ and $Y_2$ are sent through the main channel (since
in the original problem there is no side channel). 
But fortunately, since the intruder's function $\cC$ is a simple projection (in particular,
$\cC(Y_1) = Y_1 |_S$ for a small set of the coordinate positions $S \subseteq [n]$),
sending both $Y_1$ and $Y_2$ through the same channel for this special case does not 
affect the secrecy guarantees of the generalized wiretap model.
Contrary to Theorem~\ref{coro:wiretap} however, we cannot guarantee
zero leakage when we use a generalized wiretap protocol for the original
wiretap~II model.


\section*{Acknowledgment}

We would like to thank Emina Soljanin for explaining
their result \cite{ref:Emina} to us and pointing out
\cite{ref:RankMetric}.

\bibliographystyle{IEEEtran} \bibliography{full}

\appendix

\subsection{Review of the Related Notions in Cryptography}
\label{app:ERC}

In this appendix we review notions in cryptography that are relevant
to our wiretap protocol model introduced in Section~\ref{sec:model}.
These notions include \emph{resilient functions
  (RF)} and \emph{almost perfect resilient functions (APRF)},
\emph{exposure-resilient functions (ERF)}, and \emph{all-or-nothing
  transforms (AONT)}.
  
The notion of resilient functions was introduced in \cite{ref:BBR85}
(and also \cite{ref:Vaz87} as the \emph{bit-extraction problem}). A
deterministic polynomial-time computable function $f\colon \zo^n \to
\zo^m$ is called $t$-resilient \index{resilient function (RF)} if
whenever any $t$ bits of the its input are arbitrarily chosen by an
adversary and the rest of the bits are chosen uniformly at random,
then the output distribution of the function is (close to) uniform.
APRF is a stronger variation where the criterion for uniformity of the
output distribution is defined with respect to the $\ell_\infty$
(i.e., point-wise distance of distributions) rather than
$\ell_1$. This stronger requirement allows for an ``adaptive
security'' of APRFs.


ERFs\index{exposure-resilient function (ERF)}, introduced in
\cite{ref:CDH}, are similar to resilient functions except that the
entire input is chosen uniformly at random, and the view of the
adversary from the output remains (close to) uniform even after
observing any $t$ input bits of his choice.
More formally, a polynomial-time computable function $f\colon\zo^n\to\zo^m$ is
a $t$-ERF if for a uniform random variable $Y$ on $\F_2^n$ and
all $S \subseteq [n]$ such that $|S| \le t$, the two distributions
$(Y|_S, f(Y))$ and $(Y|_S, \U_m)$ are close in statistical distance.
As such, ERFs are slightly weaker than resilient functions since their
output is fully random subject to a \emph{random} fixing of positions in $S$,
as opposed to all possible fixings. A variation of ERF called \emph{adaptive ERF}
allows the adversary to make up to $t$ adaptive queries to the input; i.e.,
the choice of each query position may depend on the outcome of the previous
ones. Adaptive ERFs turn out to be much more challenging to construct than ERFs.

ERFs and resilient functions are known to be useful in a scenario
similar to the wiretap channel problem.
Suppose that Alice and Bob want to agree on any random string (for example
to use as a session key) and that their communication channel is wiretapped.
In this case Alice can send a random string $Y \in \zo^n$
to Bob and then the two parties agree on $X := f(Y)$, where $f$
is a $t$-ERF or a $t$-resilient function. It follows that an intruder who
observes any $t$ positions of $Y$ does not learn much about the agreed
string $X$. The main difference between this setting and the model we introduced
in Definition~\ref{def:wiretap} is that the two parties only care about 
the fact that $X$ is uniformly random, and not its actual realization.
That is, we have a secure \emph{agreement} problem rather than a secure \emph{communication}
problem.


Another closely related notion is that of all-or-no\-thing transforms,
which was suggested in \cite{ref:Riv97} for protection of block
ciphers.  \index{all-or-nothing transform (AONT)} This notion is defined
as follows.

\begin{defn} \label{def:AONT}
 A randomized
poly\-nomial-time computable function $f\colon \zo^m \times \zo^r \to \zo^n$, $(m
\leq n, r = \poly(n))$, where the first argument is the function's input and the second one is
a random seed, is called a (statistical, non-adaptive, and secret-only)
$t$-AONT with error $\eps$ if
\begin{enumerate}
\item It is efficiently invertible.
That is, there is a deterministic polynomial time algorithm $A$ such that
for every $x \in \zo^m$, we have $\Pr_{Z \sim \U_r}[A(f(x, Z)) = x] = 1$.
\item For
every $S \subseteq [n]$ such that $|S| \le t$, and all $x_1, x_2 \in
\zo^m$ we have that the two distributions $f(x_1,\U_r)|_S$ and $f(x_2,\U_r)|_S$
are $\eps$-close.
\end{enumerate}
\end{defn}

An AONT with $\eps=0$ is called perfect.  It is easy to see that
perfectly private wiretap protocols are equivalent to perfect adaptive
AONTs. It was shown in \cite{ref:DSS01} that such functions can not
exist (with positive, constant rate) when the adversary is allowed to
observe more than half of the encoded bits (so this is a too strong
notion for the wiretap channel problem). A similar negative result was
obtained in \cite{ref:tResilient} for the case of perfect linear RFs.

As pointed out in \cite{ref:DSS01}, AONTs can be used in the original
scenario of Ozarow and Wyner's wiretap channel problem. However, the
best known constructions of AONTs (over small alphabets) can achieve rate-resilience
trade-offs that are far from the infor\-mation-theoretic optimum (see
Figure \ref{fig:region}).
Moreover it is straightforward to see that, in the perfectly private wiretap coding
scheme of Ozarow and Wyner \cite{ref:Wyner2}, the encoder can be
seen as an adaptively secure, perfect AONTs and the decoder is in fact an
adaptive perfect RF.

While an AONT requires indistinguishability of intruder's view for
every fixed pair $(x_1, x_2)$ of messages, the relaxed notion of
\emph{average-case} AONT \index{all-or-nothing transform
  (AONT)!average case} requires the \emph{expected} distance of
$f(x_1,\U_r)|_S$ and $f(x_2,\U_r)|_S$ to be at most $\eps$ for a uniform random
message pair. That is, for average-case AONTs, the second condition in 
Definition~\ref{def:AONT} is relaxed to
\[
\Exp_{X_1, X_2 \sim \U_m}[ \dist(f(X_1,\U_r)|_S, f(X_2,\U_r)|_S ] \leq \eps.
\]
 Hence, for a negligible $\eps$, the distance will be
negligible for all but a negligible fraction of message pairs.
Here we show that, up to a loss in parameters, wiretap protocols are equivalent to
average case AONTs:

\begin{lem} \label{lem:avgAONT} Let $(E, D)$ be an encoding/decoding
  pair for a $(t, \eps, \gamma)_2$-resilient wiretap protocol. Then
  $E$ is an average-case $t$-AONT with error at most $2(\eps+\gamma)$.

  Conversely, an average-case $t$-AONT with error $\eta^4$ (for some
  small $\eta < 1/2$) can be used
  as a $(t, 2\eta, 2\eta)$-resilient wiretap encoder. 
\end{lem}

\begin{proof}
  Consider a $(t, \eps, \gamma)_2$-resilient wiretap protocol as in
  Definition~\ref{def:wiretap}, and accordingly, let the random
  variable $Y=E(X,Z)$ denote the encoding of $X$ with a random seed
  $Z$. For a set $S \subseteq [n]$ of size at most $t$, denote by $W
  := Y|_S$ the intruder's observation.

  The resiliency condition implies that, the set of bad observations
  $B_S$ has a probability mass of at most $\gamma$ and hence, the
  expected distance $\dist(X|W, X)$ taken over the distribution of $W$
  is at most $\eps+\gamma$.  Now we can apply
  Proposition~\ref{prop:duality} in the appendix to the jointly distributed pair of
  random variables $(W, X)$, and conclude that the expected distance
  $\dist(W|X, W)$ over the distribution of $X$ (which is uniform) is
  at most $\eps+\gamma$. This implies that the encoder is an
  average-case $t$-AONT with error at most $2(\eps+\gamma)$.

  For the converse, we define the
  wiretap encoder $E$ and decoder $D$ by the AONT and
  its inverse, respectively. From the definition of AONTs, it is
  immediately clear that the decodability condition of Definition~\ref{def:wiretap}
  is satisfied. In order to show the resiliency, fix any subset
  $S \subseteq [n]$ of size $t$ and assume that the intruder observes
  $W := Y|_S$, where $Y$ is the encoding of a random message $X \in \zo^m$.
  Denote by $\mathcal{W}$ the distribution of $W$ on $\zo^t$, and
  for every $x \in \zo^m$, by $\mathcal{W}_x$ the distribution of $W$
  conditioned on the event $X = x$. Now, the definition of $t$-AONT
  implies that for a random pair $X_1, X_2 \sim \U_m$, we have
  \[
  \Exp_{X_1, X_2} \dist(\mathcal{W}_{X_1}, \mathcal{W}_{X_2}) \leq \eta^4.
  \]
  Fix any $\hat{x} \in \zo^m$ for which we have
  \[
  \Exp_X \dist(\mathcal{W}_{\hat{x}}, \mathcal{W}_{X}) \leq \eta^4.
  \]
  By averaging, we know that such an $\hat{x}$ exists. By another
  averaging argument (using Markov's inequality), there is a subset $\mathcal{G} \subseteq \zo^m$ of size
  at least $(1-\eta^2)2^m$ such that
  for every $x \in \mathcal{G}$, we have
  $\dist(\mathcal{W}_{\hat{x}}, \mathcal{W}_x) \leq \eta^2$.
  Therefore, for each $x \in \mathcal{G}$, the distribution
  $\mathcal{W}_x$ (now seen as a vector of probabilities over $\zo^t$),
  can be written as
  $\mathcal{W}_x = \mathcal{W}_{\hat{x}} + \cE_x$,
  where by the definition of statistical distance, the $\ell_1$
  norm $\| \cE_x \|_1$ is at most $2\eta^2$.
 On the other hand, we can write down the convex combination
 \begin{eqnarray*}
  \mathcal{W} &=& 2^{-m} \sum_{x \in \zo^m} \mathcal{W}_x \\
  &=& 2^{-m} \sum_{x \in \mathcal{G}} \mathcal{W}_x + 2^{-m} \sum_{x \in \zo^m \setminus \mathcal{G}} \mathcal{W}_x\\
  &=& 2^{-m} \sum_{x \in \mathcal{G}} \mathcal{W}_x + \cE,
 \end{eqnarray*}
 where we have defined $\cE := 2^{-m} \sum_{x \in \zo^m \setminus \mathcal{G}} \mathcal{W}_x$, and know that
 $\| \cE \|_1 \leq \eta^2$.  Substituting for $\mathcal{W}_x$, we get
 \begin{eqnarray*}
  \mathcal{W} 
  &=& 2^{-m} \sum_{x \in \mathcal{G}} (\mathcal{W}_{\hat{x}} + \cE_x) + \cE \\
  &=& 2^{-m} |\mathcal{G}| \mathcal{W}_{\hat{x}} + \cE',
 \end{eqnarray*}
 where $\cE' := 2^{-m} \sum_{x \in \mathcal{G}}\cE_x + \cE$. By the
 bounds on the $\ell_1$ norm on the $\cE_x$, we know that 
 $\|\cE'\|_1 \leq 3 \eta^2$, which implies
 \begin{equation} \label{eqn:WsumWx}
 \| \mathcal{W} - 2^{-m} |\mathcal{G}| \mathcal{W}_{\hat{x}} \|_1 \leq 3\eta^2. 
\end{equation} Now, since $2^{-m} |G| \geq 1-\eta^2$, 
 we have
 \[
 \| \mathcal{W} - \mathcal{W}_{\hat{x}} \|_1 \leq
 \| \mathcal{W} - 2^{-m} |\mathcal{G}| \mathcal{W}_{\hat{x}} \|_1 + \eta^2
 \leq 4\eta^2,
 \]
 where the second inequality is from \eqref{eqn:WsumWx}. So, the distributions $\mathcal{W}$
 and $\mathcal{W}_{\hat{x}}$ are $(2\eta^2)$-close.
 Now we can apply Proposition~\ref{prop:duality} for the joint distribution of 
 the variables $(X,W)$, and deduce
 \begin{equation} \label{eqn:distXWUm}
 \Exp_W[\dist(X|W, \U_m)] = \Exp_X[\dist(\mathcal{W}_X, \mathcal{W})].
 \end{equation}
 Here we have used the notation $X|W$ for the distribution of the random variable
 $X$ conditioned on the observation $W$.
 We know that for every $x \in \mathcal{G}$, 
 \[
 \dist(\mathcal{W}_x, \mathcal{W}) \leq \dist(\mathcal{W}_x, \mathcal{W}_{\hat{x}}) + 2\eta^2 \leq 3 \eta^2.
 \] 
 Thus,
 \[
 \Exp_X[\dist(\mathcal{W}_X, \mathcal{W})] \leq |\zo^m \setminus \mathcal{G}| 2^{-m} + 3 \eta^2 \leq 4 \eta^2. 
 \]
 Combined with \eqref{eqn:distXWUm}, we get
 \[
 \Exp_W[\dist(X|W, \U_m)] \leq 4 \eta^2.
 \]
 Now we are ready to apply Markov's inequality again to conclude that
 \[
 \Pr[\dist(X|W, \U_m) > 2\eta] \leq 2 \eta,
 \]
 which proves the resiliency condition.
\end{proof}

It is important to note that the converse direction does not guarantee zero leakage,
and hence, zero leakage wiretap protocols are in general stronger than
average-case AONTs.  An average-case to worst-case reduction for AONTs
was shown in \cite{ref:CDH} which, combined with the above lemma, can
be used to show that any wiretap protocol can be used to construct an
AONT (at the cost of a rate loss).

A simple \emph{universal} transformation was proposed in
\cite{ref:CDH} to obtain an AONT from any ERF, by one-time padding the
message with a random string obtained from the ERF.  In particular,
given an ERF $f\colon \zo^n \to \zo^m$, the AONT $g\colon \zo^m \times \zo^n \to
\zo^{m+n}$ is defined as $g(x) := (z,x + f(z))$.  
In other words, the ERF is used to one-time pad
the message with a random secret string. 
As proved in in \cite{ref:CDH}, the function $g$ constructed in this
way is indeed an AONT.

This construction can also yield a wiretap protocol with zero leakage
(this is easy to see by following the proof of \cite{ref:CDH} for the
above construction).
However, it has the drawback of significantly weakening the
rate-resilience trade-off. Namely, even if an information
theoretically optimal ERF is used in this reduction, the resulting
wiretap protocol will only achieve half the optimal rate (see
Fig.~\ref{fig:region}).  This is because the one-time padding
strategy necessarily requires a random seed that is at least as long
as the message itself, even if the intruder is restricted to observe
only a small fraction of the transmitted sequence. Hence the rate of
the resulting AONT cannot exceed $1/2$, and it is not clear how to
improve this universal transformation to obtain a worst-case AONT
using a shorter seed.

For applications in cryptography, e.g., the
context of ERFs or AONTs, it is typically assumed that the adversary
learns \emph{all} but a small number of the bits in the encoded
sequence, and the incurred \emph{blow-up} in the encoding is not as
crucially important, as long as it remains within a reasonable
range. On the other hand, as in this work we are motivated by the
wiretap channel problem which is a communication problem, optimizing
the transmission \emph{rate} will be the most important concern for
us.

  \subsection{Some Technical Details} \label{app:technical}

  In this appendix we present some minor technical details and tools that are
  omitted in the main text.

  The following proposition quantifies the Shannon entropy of a
  distribution that is close to uniform:

  \begin{prop} \label{prop:Shannon} Let $\cX$ be a probability
    distribution on a finite set $S$, $|S| > 4$, that is $\eps$-close
    to the uniform distribution on $S$, for some $\eps \leq 1/4$. Then
    $H(\cX) \geq \log_2 |S| (1 - \eps)
    $ 
  \end{prop}

\begin{Proof}
  Let $n := |S|$, and let $f(x) := - x \log_2 x$. The function $f(x)$
  is concave, passes through the origin and is strictly increasing in
  the range $[0, 1/\mathrm{e}]$.  From the definition, we have $H(\cX)
  = \sum_{s \in S} f(\Pr_\cX(s))$.  For each term $s$ in this
  summation, the probability that $\cX$ assigns to $s$ is either at
  least $1/n$, which makes the corresponding term at least $\log_2
  n/n$ (due to the particular range of $|S|$ and $\eps$), or is equal
  to $1/n - \eps_s$, for some $\eps_s > 0$, in which case the term
  corresponding to $s$ is less than $\log_2 n/n$ by at most $\eps_s
  \log_2 n$ (this follows by observing that the slope of the line
  connecting the origin to the point $(1/n, f(1/n))$ is $\log_2
  n$). The bound on the statistical distance implies that the
  differences $\eps_s$ add up to at most $\eps$.  Hence, the Shannon
  entropy of $\cX$ can be less than $\log_2 n$ by at most $\eps \log_2
  n $.
\end{Proof}

\begin{prop} \label{prop:duality} Let $(X, Y)$ be a pair of random
  variables jointly distributed on a finite set $\Omega \times
  \Gamma$. Then\footnote{Here we are abusing the notation and denote
    by $Y$ the marginal distribution of the random variable $Y$, and
    by $Y|(X=a)$ the distribution of the random variable $Y$
    conditioned on the event $X=a$.}  $ \Exp_Y[\dist(X|Y, X)] =
  \Exp_X[\dist(Y|X, Y)].  $
\end{prop}

\begin{Proof}
  For $x \in \Omega$ and $y \in \Gamma$, we will use shorthands $p_x,
  p_y, p_{xy}$ to denote $\Pr[X=x], \Pr[Y=y],$ and $\Pr[X=x,Y=y]$,
  respectively. Then we have
  \begin{eqnarray*}
    \Exp_Y[\dist(X|Y, X)] &=& \sum_{y \in \Gamma} p_y \dist(X|(Y=y), X)
    \\ &=& \frac{1}{2} \sum_{y \in \Gamma} p_y \sum_{x \in \Omega} |p_{xy}/p_y - p_x| \\
    &=& \frac{1}{2} \sum_{y \in \Gamma} \sum_{x \in \Omega} |p_{xy} - p_x p_y| 
    \\ &=& \frac{1}{2} \sum_{x \in \Omega} p_x \sum_{y \in \Gamma} |p_{xy}/p_x - p_y| \\
    &=& \sum_{x \in \Omega} p_x \dist(Y|(X=x), Y) \\ &=& \Exp_X[\dist(Y|X, Y)].
  \end{eqnarray*}
\end{Proof}

\begin{prop} \label{prop:conditioning} Let $\Omega$ be a finite set
  that is partitioned into subsets $S_1, \ldots, S_k$ and suppose that
  $\cX$ is a distribution on $\Omega$ that is $\gamma$-close to
  uniform. Denote by $p_i$, $i = 1, \ldots k$, the probability
  assigned to the event $S_i$ by $\cX$. Then
  \[
  \sum_{i \in [k]} p_i \cdot \dist( \cX | S_i, \U_{S_i} ) \leq 2
  \gamma.
  \]
\end{prop}

\begin{Proof}
  Let $N := |\Omega|$, and define for each $i$, $ \gamma_i := \sum_{s
    \in S_i} \left| \Pr_\cX(s) - \frac{1}{N} \right|,$ so that
  $\gamma_1 + \cdots + \gamma_k \leq 2 \gamma$.  Observe that by
  triangle's inequality, for every $i$ we must have $|p_i - |S_i|/N |
  \leq \gamma_i$.  To conclude the claim, it is enough to show that
  for every $i$, we have $\dist( \cX | S_i, \U_{S_i} ) \leq \gamma_i /
  p_i$.  This is shown in the following.
  {\allowdisplaybreaks \begin{eqnarray*}
   && p_i \cdot \dist( \cX | S_i, \U_{S_i} ) = \frac{p_i}{2} \sum_{s \in S_i} \left| \frac{\Pr_\cX(s)}{p_i} - \frac{1}{|S_i|} \right| \\
    &=& \frac{1}{2} \sum_{s \in S_i} \left| \Pr_\cX(s) - \frac{p_i}{|S_i|} \right| \\
    &=& \frac{1}{2} \sum_{s \in S_i} \left| \left(\Pr_\cX(s) - \frac{1}{N} \right) + \frac{1}{|S_i|} \left( \frac{|S_i|}{N} - p_i \right) \right| \\
    &\leq& \frac{1}{2} \sum_{s \in S_i} \left| \Pr_\cX(s) - \frac{1}{N} \right| + \frac{1}{2|S_i|} \sum_{s \in S_i} \left| \frac{|S_i|}{N} - p_i  \right| \\
    &\leq& \frac{\gamma_i}{2} + \frac{1}{2|S_i|} \cdot |S_i| \gamma_i = \gamma_i. 
  \end{eqnarray*}}
\end{Proof}

The following proposition shows that any function
maps close distributions to close distributions:

\begin{prop} \label{prop:closeFunction} Let $\Omega$ and $\Gamma$ be
  finite sets and $f$ be a function from $\Omega$ to $\Gamma$. Suppose
  that $\mathcal{X}$ and $\mathcal{Y}$ are probability distributions
  on $\Omega$ and $\Gamma$, respectively, and let $\mathcal{X'}$ be a
  probability distribution on $\Omega$ which is $\delta$-close to
  $\mathcal{X}$. Then if $f(\mathcal{X}) \sim_\eps \mathcal{Y}$, then
  $f(\mathcal{X'}) \sim_{\eps+\delta} \mathcal{Y}$. 
\end{prop}

\begin{proof}
Let $X$, $X'$ and $Y$ be random variables distributed
according to $\mathcal{X}$, $\mathcal{X'}$, and $\mathcal{Y}$,
respectively. We want to upperbound
\[
 \left| \Pr[ f(X') \in T ] - \Pr[ Y \in T ] \right|
\]
for every $T \subseteq \Gamma$.
By the triangle inequality, this is no more than
\begin{multline*}
 \left| \Pr[ f(X') \in T ] - \Pr[ f(X) \in T ] \right| + \\ \left| \Pr[ f(X) \in T ] - \Pr[ Y \in T ] \right|.
\end{multline*}
Here the summand on the right hand side is upperbounded by the distance
of $f(\mathcal{X})$ and $\mathcal{Y}$, that is assumed to be at most $\eps$.
Let $T' \eqdef \{x \in \Omega \mid f(x) \in T\}$. 
Then the summand on the left can be written as
\[
 \left| \Pr[X' \in T'] - \Pr[X \in T'] \right| 
\]
which is at most $\delta$ by the assumption that $\mathcal{X} \sim_\delta \mathcal{X'}$.
\end{proof}

\subsection{Omitted Details of the Proof of
  Corollary~\ref{coro:WTWalk}} \label{app:wiretapDetails}

\newcommand{\MOD}{\mathrm{mod}\ } \newcommand{\modf}{\mathsf{Mod}}
\newcommand{\imod}{\mathsf{Inv}}

Here we prove Corollary~\ref{coro:WTWalk} for the case $c > 1$.  The
construction is similar to the case $c = 1$, and in particular the
choice of $m$ and $k$ will remain the same. However, a subtle
complication is that the expander family may not have a graph with
$d^m$ vertices and we need to adapt the extractor of
Theorem~\ref{thm:invWalk} to support our parameters, still with
exponentially small error. To do so, we pick a graph $G$ in the family
with $N$ vertices, such that \[c^{\eta m} d^m \leq N \leq c^{\eta m +
  1} d^m,\] for a small absolute constant $\eta > 0$ that we are free
to choose.  The assumption on the expander family guarantees that such
a graph exists. Let $m'$ be the smallest integer such that $d^{m'}
\geq c^{\eta m} N$.  Index the vertices of $G$ by integers in $[N]$.
Note that $m'$ will be larger than $m$ by a constant multiplicative
factor that approaches $1$ as $\eta \to 0$.

For positive integers $q$ and $p \leq q$, define the function
$\modf_{q, p}\colon [q] \to [p]$ by \[\modf_{q, p}(x) := 1 + (x\ \MOD
p).\] The extractor $\kz$ interprets the first $m'$ symbols of the
input as an integer $v$, $0\le v< d^{m'}$ and performs a walk on $G$
starting from the vertex $\modf_{d^{m'}, N}(u+1)$, the walk being
defined by the remaining input symbols. If the walk reaches a vertex
$u$ at the end, the extractor outputs $\modf_{N, d^{m}}(u)-1$, encoded
as a $d$-ary string of length $m$.  A similar argument as in
Theorem~\ref{thm:invWalk} can show that with our choice of the
parameters, the extractor has an exponentially small error, where the
error exponent is now inferior to that of Theorem~\ref{thm:invWalk} by
$O(m)$, but the constant behind $O(\cdot)$ can be made arbitrarily
small by choosing a sufficiently small $\eta$.

The real difficulty lies with the inverter because $\modf$ is not a
balanced function (that is, all images do not have the same number of
preimages), thus we will not be able to obtain a perfect
inverter. Nevertheless, it is possible to construct an inverter with a
close-to-uniform output in $\ell_\infty$ norm. This turns out to be as
good as having a perfect inverter, and thanks to the Lemma~\ref{lem:infty} below,
we will still be able to use it to construct a wiretap protocol with
zero leakage. Intuitively, the reason is that even though
we are not working with a perfect inverter and thus the distribution
of the encoding conditioned on the intruder's observation will not
precisely be a symbol-fixing source, but it is extremely close to one,
and the proof of Theorem~\ref{thm:protocol} for the case of zero
leakage will still go through.

\begin{lem}
  \label{lem:infty}
  Suppose that $f\colon [d]^n \rightarrow [d]^m $ is a
  $(k,2^{-\Omega(m)})_d$ symbol-fixing extractor and that $\cX$ is a
  distribution on $[d]^n$ such that $\|\mathcal{X}-\U_{[d]^n}\|_\infty
  \le 2^{-\Omega(m)}/d^n$. Denote by $\cX'$ the distribution $\cX$
  conditioned on any fixing of at most $n-k$ coordinates. Then
  $f(\cX') \sim_{2^{-\Omega(m)}} \U_{[d]^m}$.
\end{lem}
\begin{Proof}
  By Proposition~\ref{prop:closeFunction}, it suffices to show that
  $\cX'$ is $2^{-\Omega(m)}$-close to an $(n, k)_d$ symbol-fixing
  source.  Let $S \subseteq [d]^m$ denote the support of $\cX'$, and
  let $\eps/d^n$ be the $\ell_\infty$ distance between $\mathcal{X}$
  and $\U_{[d]^n}$, so that by our assumption, $\eps =
  2^{-\Omega(m)}$.  By the bound on the $\ell_\infty$ distance, we
  know that $\Pr_\mathcal{X}(S)$ is between $\frac{|S|}{d^n}(1 -
  \eps)$ and $\frac{|S|}{d^n}(1 + \eps)$. Hence for any $x \in S$,
  $\Pr_{\cX'}(x)$, which is $\Pr_{\cX}(x)/\Pr_{\cX}(S)$, is between
  $\frac{1}{|S|}\cdot \frac{1 - \eps}{1 + \eps}$ and
  $\frac{1}{|S|}\cdot \frac{1 + \eps}{1 - \eps}$.  This differs from
  $1/|S|$ by at most $O(\eps)/|S|$. Hence, $\cX'$ is
  $2^{-\Omega(m)}$-close to $\U_S$.
\end{Proof}

In order to invert our new construction, we will need to construct an
inverter $\imod_{q, p}$ for the function $\modf_{q, p}$.  For that,
given $x \in[p]$ we will just sample uniformly in its preimages.  This
is where the non-balancedness of $\modf$ causes problems, since if $p$
does not divide $q$ the distribution $\imod_{q, p}(\U_{[p]})$ is not
uniform on $[q]$.
\begin{lem}
  Suppose that $q>p$.  Given a distribution $\mathcal{X}$ on $[p]$
  such that $\| \mathcal{X} - \U_{[p]} \|_\infty \le \frac{\eps}{p}$,
  we have $\| \imod_{q, p}(\mathcal{X}) - \U_{[q]} \|_\infty \le
  \frac{1}{q}\cdot \frac{p+\epsilon q}{q-p}$.
\end{lem}
\begin{Proof}
  Let $X \sim \mathcal{X}$ and $Y \sim \imod_{q,p}(\mathcal{X})$.
  Since we invert the modulo function by taking for a given output a
  random preimage uniformly, $\Pr[Y=y]$ is equal to $\Pr[X =
  \modf_{q,p}(y)]$ divided by the number of $y$ with the same value
  for $\modf_{q,p}(y)$.  The latter number is either $\lfloor q/p
  \rfloor$ or $\lceil q/p \rceil$, so
  \begin{equation*}
    \frac{1-\eps}{p \lceil q/p \rceil} \le \Pr(Y = y) \le 
    \frac{1+\eps}{p \lfloor q/p \rfloor}
  \end{equation*}
  Bounding the floor and ceiling functions by $q/p \pm 1$, we obtain
  \begin{equation*}
    \frac{1-\eps}{q+p}
    \le \Pr(Y = y) \le 
    \frac{1+\eps}{q-p}
  \end{equation*}
  That is
  \begin{equation*}
    \frac{-p-\epsilon q}{q(q+p)} 
    \le \Pr(Y = y)-\frac{1}{q} \le 
    \frac{p+\epsilon q}{q(q-p)}\ ,
  \end{equation*}
  which concludes the proof since this is true for all $y$.
\end{Proof}

Now we describe the inverter $\inv(x)$ for the extractor.
  First the inverter calls $\imod_{N, d^{m}}(x)$
to obtain $x_1 \in [N]$.  Then it performs a random walk of length $n-m'$
on the graph,
starting from $x_1$, to reach a vertex $x_2$ at the end which is
inverted to obtain $x_3=\imod_{d^{m'}, N}(x_2)$ as a $d$-ary string of
length $m'$.  Finally, the inverter outputs $y=(x_3,w)$, where $w$
corresponds the \emph{inverse} of the random walk.
It is obvious that this procedure yields a valid preimage of $x$.

Using the previous lemma, if $x$ is chosen uniformly, $x_1$ will be at
$\ell_\infty$-distance \[\epsilon_1 := \frac{1}{N}\cdot
\frac{d^m}{N-d^m} = \frac{1}{N}O(c^{-\eta m}).\] For a given walk, the
distribution of $x_2$ will just be a permutation of the distribution
of $x_1$ and applying the lemma again, we see that the
$\ell_\infty$-distance of $x_3$ from the uniform distribution is
\[ \epsilon_2 := \frac{1}{d^{m'}} \cdot \frac{N+\epsilon_1
  d^{m'}}{d^{m'}-N} = \frac{1}{d^{m'}} O(c^{-\eta m}).\] This is true
for all the $d^{n-m'}$ possible walks so the $\ell_\infty$-distance of
the distribution of $y$ from uniform is bounded by $\frac{1}{d^n}
O(c^{-\eta m})$.  Applying Lemma~\ref{lem:infty} in an argument
similar to Theorem~\ref{thm:protocol} concludes the proof.  

\end{document}